\documentclass{ws-jai}
\usepackage{epstopdf}
\usepackage{subfig}
\begin{document}

\catchline{}{}{}{}{} 

\markboth{Alessio Magro}{Mulitbeam GPU Transient Pipeline for the Medicina BEST-2 Array}

\title{Mulitbeam GPU Transient Pipeline for the Medicina BEST-2 Array}

\author{Alessio Magro$^1$, Jack Hickish$^2$ and Kristian Zarb Adami$^1$}

\address{
$^1$Department of Physics, University of Malta, Msida, MSD 2080, Malta \\
$^2$Department of Physics, University of Oxford, Denys Wilkinson Building, Keble Road, Oxford OX1 3RH. UK
}

\maketitle

\begin{history}
\received{(to be inserted by publisher)};
\revised{(to be inserted by publisher)};
\accepted{(to be inserted by publisher)}; 
\end{history}

\begin{abstract}
Radio transient discovery using next generation radio telescopes will pose several digital signal processing and data
transfer challenges, requiring specialized high-performance backends. Several accelerator technologies are being considered
as prototyping platforms, including Graphics Processing Units (GPUs). In this paper we present a real-time pipeline
prototype capable of processing multiple beams concurrently, performing Radio Frequency Interference (RFI) rejection
through thresholding, correcting for the delay in signal arrival times across the frequency band using brute-force dedispersion, 
event detection and clustering, and finally candidate filtering, with the capability of persisting data buffers containing 
interesting signals to disk. This setup was deployed at the BEST-2 SKA pathfinder in Medicina, Italy, where several 
benchmarks and test observations of astrophysical transients were conducted. These tests show that on the deployed hardware	
eight 20MHz beams can be processed simultaneously for $\sim$640 Dispersion Measure (DM) values. Furthermore, the clustering
and candidate filtering algorithms employed prove to be good candidates for online event detection techniques. The number
of beams which can be processed increases proportionally to the number of servers deployed and number of GPUs, making
it a viable architecture for current and future radio telescopes.
\end{abstract}


\section{Introduction}

One of the key science projects of several next-generation radio telescopes is to explore the nature of the dynamic radio sky with ever increasing spectral 
and temporal resolution. The speed at which blind surveys for radio transients can be performed is on the increase as well,
with multiple beams capable of scanning different parts of the sky in parallel. Such surveys generate large amounts of data,
in the order of gigabtyes per second,
which makes the notion of saving all this incoming data for future processing unfeasible. Real-time systems have to be 
employed to filter this data and only save segments containing potentially interesting events. Such systems also offer the 
possibility of reacting to such events in real-time, possibly taking control of the telescope's monitoring system and perform
tracking observations using a subset of the beams. The idea of triggering other telescopes operating across
the entire electromagnetic spectrum and performing joint follow-up observations is also an attractive and realizable one. 

These concepts have been among the driving forces for several real-time transient detection prototypes, such as the 
GMRT digital backend which is implemented on standard, commercial, off-the-shelf components \cite{roy10,bhat13}. Due to the high
computational requirement for blind transient surveys, alternative hardware architectures are also being investigated, with significant
focus on Graphics Processing Units (GPUs), which are the main components of several prototypes being 
developed, such as the ones at the Parkes Radio Telescope \cite{barsdell12} and Low Frequency Array (LOFAR) single stations \cite{armour12}. 
In this work we expand on the system
developed in \cite{magro11} and transform it into a standalone, scalable, high-throughput transient detection system. We 
consider the case where beamformed data is processed to extract astrophysical radio bursts of short duration. \cite{cordes03}
have thoroughly discussed the range of potential events which can produce such transients, ranging from individual 
neutron star emissions to extragalctic millisecond bursts \cite{lorimer07,keane2012}. We deploy this system on the BEST-2 array in Medicina, where filterbank data is 
received from  an FPGA-based digital beamformer implemented on 
ROACH\footnote{Reconfigurable Open Architecture Computing Hardware - https://casper.berkeley.edu/wiki/ROACH}-boards

Apart from the high computational requirements for an online transient detection pipeline, there are also two major issues which 
need to be tackled: mitigation of signals induced by terrestrial radio sources, and an accurate event detection mechanism which 
reduces the data output of the system whilst minimising as much as possible the detection of false positives. Depending on the 
degree and type of Radio Frequency Interference (RFI) different mitigation techniques might need to be used. We choose to implement
a simple thresholding mechanism that removes any spectra or frequency slices which exceed a certain threshold. In order to avoid
the risk of thresholding high-power dispersed radio pulses the algorithm will allow low power RFI to seep through. These events will
then be classified in the detection stage, which first clusters data points in {\it dm-snr-time} space and then applies a filter
to generate transient candidates and remove clusters caused by RFI events. This filter will compare each cluster's {\it dm-snr}
signature with one generated analytically and will classify them depending on the error difference between the two.

This paper is organized as follows: In section \ref{best2Array} we give an overview of the BEST-2 Array and the digital backend
In section \ref{transientPipeline} we describe in some detail the implementation of our GPU-based 
processing pipeline, while in section \ref{benchmarks} we benchmark the entire pipeline. In section
\ref{deployment} we describe the deployment setup as well as some initial test observation conducted using the system, and
in section \ref{conclusion} we present our conclusions.

\subsection{BEST-2 Array and Digital Backend}
\label{best2Array}

The BEST-2 \cite{montebugnoli09} testedbed at the Radiotelescopi di Medicina, located near Bologna, Italy, is composed of eight East-West
oriented cylindrical concentrators each having 64-dipole receivers critically sampled at 408 MHz with a bandwidth
of 16 MHz. Signals from these dipoles are combined in groups of 16 using analog circuitry, resulting in four analog 
channels
per cylinder, providing  a total of 32 effective elements positioned on a 4 x 8 grid. These signals are fed to 
the ROACH-based digital backend developed by \cite{hickish13} and \cite{foster12}, where they are digitized and 
channelized into a total of 1024 single-polarization frequency channels. The digital backend processed 20 MHz of
bandwidth, even though only 16 MHz are useful.
A spatial fourier transform is then 
performed to create 128 beams, after which 8 beams are selected for output as 16-bit complex values over a 10GbE 
interface using a custom SPEAD\footnote{https://casper.berkeley.edu/wiki/SPEAD} packet format.
The SPEAD protocol aims to standardise the UDP datastream format as output by radio astronomy instruments,
defining the datastream as self describing, meaning that transmit and receive code can be used for multiple
telescopes since the layout of the packets are well defined. The output rate of a single beam from the
beamformer is 640 Mbps excluding packet headers, which is calculated using $D = C \times T \times W$, 
where C is the number of frequency channels, $T$ is the number of time samples per second and $W$ is the word length. 
In our case, $C$ = 1024, $T$ = 19531.25 (20 MHz processable bandwidth channelized into 1024 channels) and $W$ = 32 bits 
(16 bits for each complex component). A total output bandwidth of 5.12 Gbps is required to send out all 8 beams, which is
manageable over a single 10GigE link. In this paper we will only concern ourselves with the technical specification of 
this setup and how the outgoing data can be processed on a single processing server. 

\section{Transient Detection Pipeline}
\label{transientPipeline}

\begin{figure}[t]
\begin{center}
\includegraphics[width=420pt]{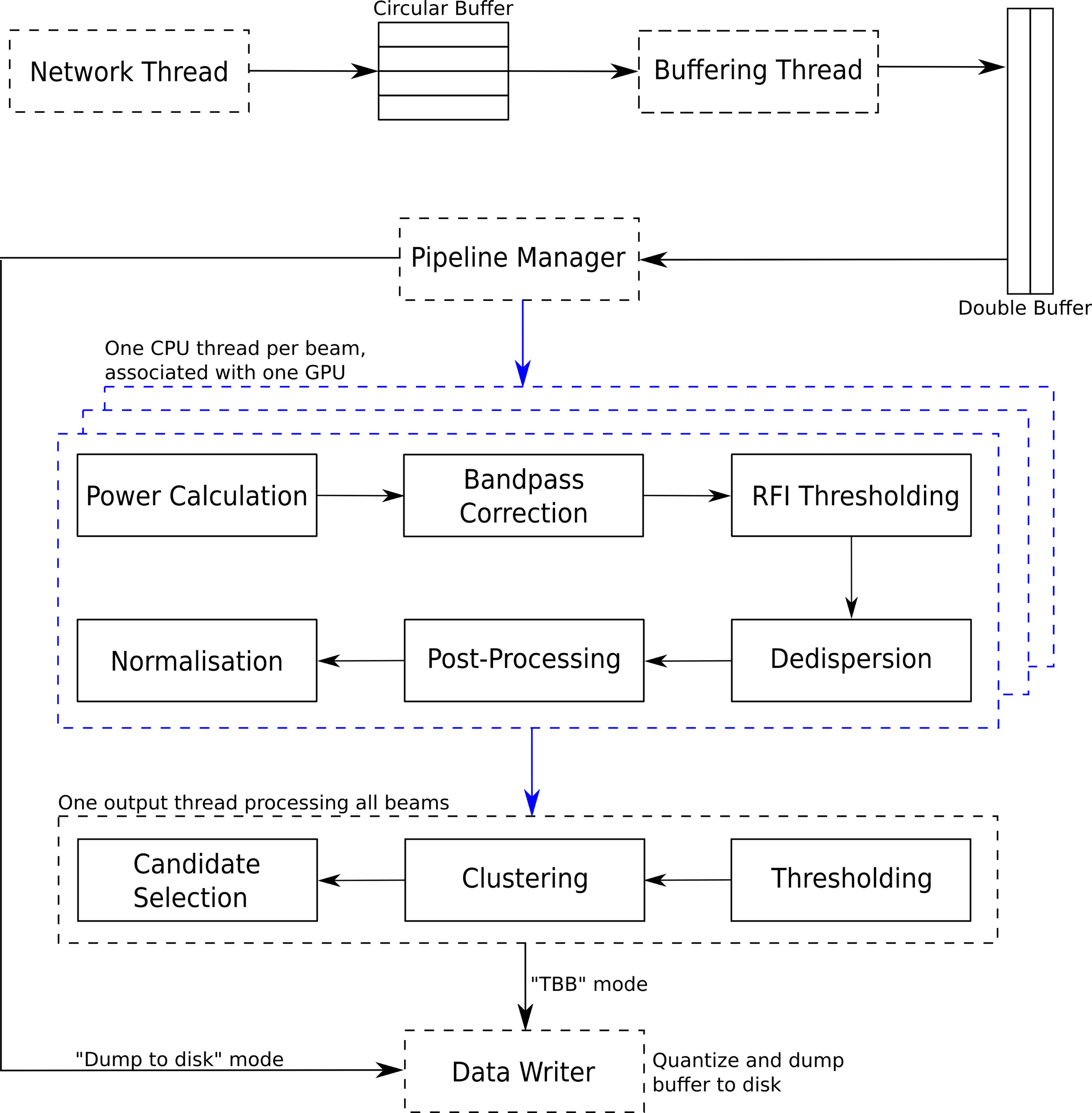}
\end{center}
\caption{Overview of the system and pipeline architecture. At the receiving end two threads take care
	 of packet reception, interpretation and buffering. These populate a double buffer, which is read
	 by the Pipeline Manager and copied to GPU memory where it is processed. A CPU thread is created
	 for each beam and is associated with one GPU (which may be shared with other threads). 
	 The processed data is then copied back to host memory where it is thresholded, clustered and 
	 filtered using a separate lightweight thread per beam. During the entire execution data buffers can
	 be persisted to disk by sending signals to the Data Writer thread. Boxes in dashed lines represent
	 distinct CPU threads, whilst dashed blue overlays represent CPU threads which use GPUs for processing.
	 Blue arrows represent data transfers to or from GPU memory. For more details see section 
	 \ref{transientPipeline}.}
\label{architectureFigure}
\end{figure}

We have designed and implemented a real-time, GPU-based, transient detection pipeline for the BEST-2 array. We 
extended the pipeline design adopted in \cite{magro11} to include a fast buffering system, the ability to
write streamed data to disk at different stages in the pipeline, the ability to process multiple beams concurrently 
and an online candidate selection mechanism to separate RFI events from astrophysical ones. 
The main design emphasis was high-performance and scalability across many beams. The high-level architecture of 
this pipeline is depicted in figure \ref{architectureFigure}, which is split into three main processing stages: 
the data reception and buffering stage, the GPU-based processing pipeline stage and the post-processing stage.
Each beam is processed on a single GPU, having an associated CPU tread, where a GPU is capable of processing 
multiple beams depending on the amount of processing required.. The current implementation assumes that beams are 
identical (same observation and survey 
parameters), however this can be extended to allow the possibility of performing different operations on each beam.

Packet reception, interpretation and buffering is performed on the CPU, which then forwards the data to the GPU
where it is passes through several processing stages including: power calculation, bandpass correction, RFI thresholding, dedispersion
and optional post-processing and normalization, after which the dedispersed time-series are copied back to
CPU memory and passed through a detection stage where it is thresholded, clustered and classified. Any data-points belonging
to interesting clusters (having a high probability that they're not due to RFI events) are written to disk, together
with the unprocessed data buffer after being quantized to 8 or 4 bits depending on whether the data has been converted
to a channelized power-series in the receiver thread. There is also the possibility of writing the entire data stream
to disk, including buffers without interesting events, after passing through an encoding and quantization stage, 
provided that the disk drives can manage the data rate. 
All operating parameters are provided by an XML configuration file.

\subsection{Data Receiver and Buffering}

\begin{figure}[t]
\begin{center}
\includegraphics[width=450pt]{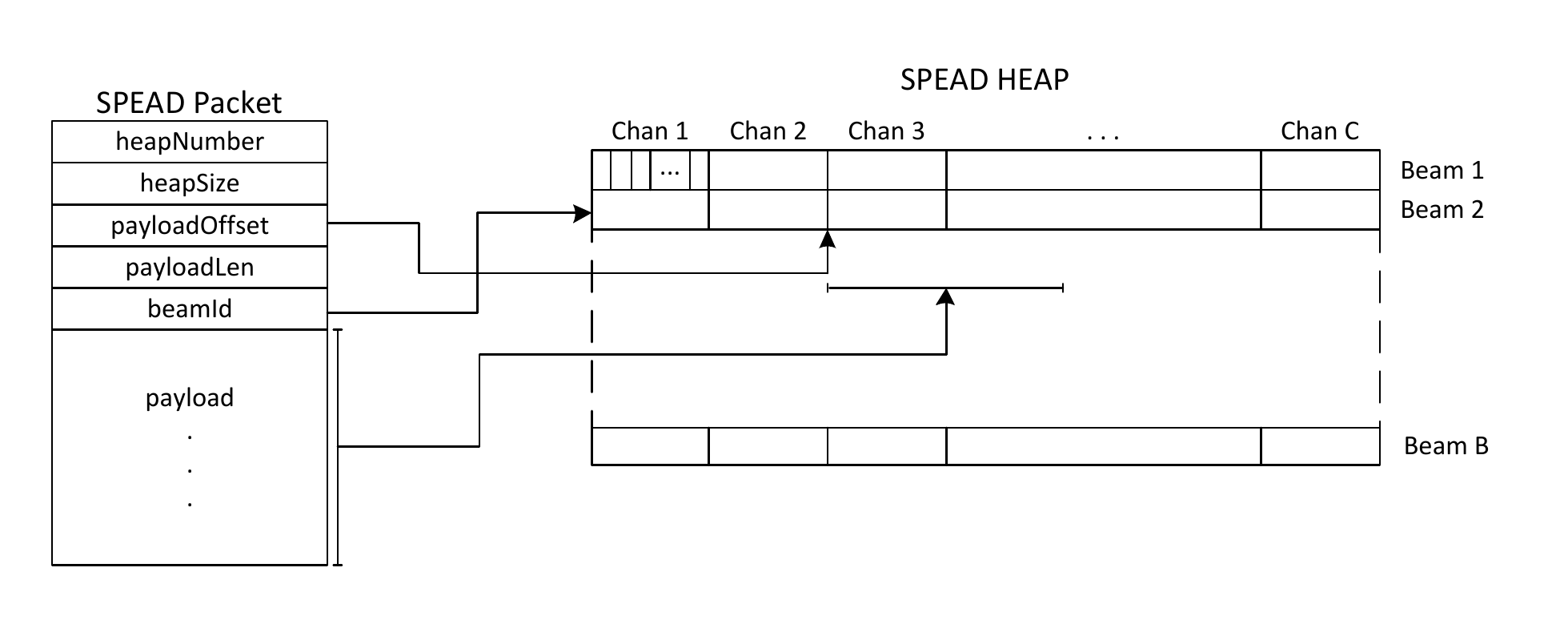}
\end{center}
\caption{SPEAD packet format and heap data organization used to send data between the beamformer and GPU server. A 
heap contains a number of spectra for all the beams being processed, for all the band. This is split up across multiple 
packets which contain time samples from a subset of the frequency channels, for a single beam. The packet header defines
where the data should be located within the heap, as well as additional information to be able to identify heaps and
extract timestamp information.}
\label{packetFormatFigure}
\end{figure}

The Medicina beamformer packetizes the channelized beamformed data using a custom SPEAD
packet format designed to reduce the overhead of heap generation and data movement on the receiver side.
A heap consists of a time-slice containing several spectra (composed of 16-bit complex values sampled 
for all channels) from multiple beams, organized in beam/channel/time order which maps directly to the 
data organization in GPU memory, thus considerably reducing the overhead for memory re-arrangement. 
Figure \ref{packetFormatFigure} describes in further detail the data organization of a heap and how this 
maps to the packet format. The UDP transmission protocol is used to send the packet stream since it is 
lightweight and the probability of losing packets, or receiving them out of order, is very small and can 
be easily handled using simple error checking routines on the receiver side where heap buffers 
provide a small time window during which an out-of-order packet can still be processed correctly. The
heap size is set by the digital backend.

The incoming UDP stream is received and buffered for processing using two CPU threads, the {\it network thread} and the 
{\it buffering thread}. The network thread is responsible for reading and interpreting UDP packets, and its 
performance is limited by the packet size and number of operations required per packet. The latter is alleviated by
using raw sockets and allocating a circular frame buffer in kernel memory which is mapped to the process' user memory. When a 
new Ethernet frame is received, the kernel notifies the busy waiting network thread by changing appropriate values in 
this memory space, thus drastically reducing the number of system calls required. A 4 KB packet size is used to
further reduce overhead costs. The received frame is then stripped from several protocol headers and the underlying
SPEAD packet is extracted. The SPEAD header defines the heap it belongs to and where its payload should be placed 
within the heap, after which it is copied to the specified offset. This requires a heap buffer which is kept local
to the network thread until the entire heap has been received. A circular heap buffer is shared between the network and
buffering threads, allowing their operations to overlap and minimize locking overheads. When a heap is fully read, the
buffering thread is notified and a new heap buffer is returned to the network thread, which starts receiving the 
next heap. If a packet from the next heap is received before the current heap is fully populated, then all missing
packets are considered dropped, thus resulting in zeroed-out ``gaps'' in the heap, which are then handled by the RFI thresholding
stage. Increasing the number of slots in the circular buffer reduces the probability that the network thread
is kept waiting for the buffering thread to free up slots, which might happen when CPU-intensive tasks are scheduled
on the same core as the buffering thread.

The buffering thread's main function is to create larger data buffers to be copied directly to GPU memory, composed
of multiple heaps. To overlap the creation of these buffers with GPU execution, a double-buffering system is used.
Heaps from the circular buffer are separated into chunks containing a time-slice from a single channel per beam,
and these are copied to their respective locations within the GPU buffer. When this is fully populated, the main pipeline
thread is notified and the pipeline is advanced by one iteration after all the GPU-based processing has 
finished for the previous one.

\subsection{GPU Processing Pipeline}

The main pipeline consists of several processing modules which run on the GPU, with a single managing thread taking
care of initialization and synchronization. Execution is parallelized across beams, where a separate processing 
thread is initialized for each beam and is associated with a particular GPU. Each processing thread takes care of 
copying its input data from CPU to GPU memory and vice versa, which maximizes the PCIe-bandwidth used during these
operations on multi-GPU systems. The CPU imprint of these threads is almost negligible and mostly consists of kernel timing and 
synchronization, and a one-off $p^{th}$-order polynomial fit over $N_{chans}$ values per iteration. The processing
threads' execution passes through several kernel launches which perform specific functions. Although we could get
a potential speedup by combining some of these kernels into monolithic functions, separating functionality this
way makes it easier to add intermediary stages as well as disabling certain functionality when required. 

\subsubsection{RFI Thresholding}

\begin{figure}
  \centering 
  \subfloat[Broadband RFI spike with no thresholding]{\includegraphics[width=200pt]{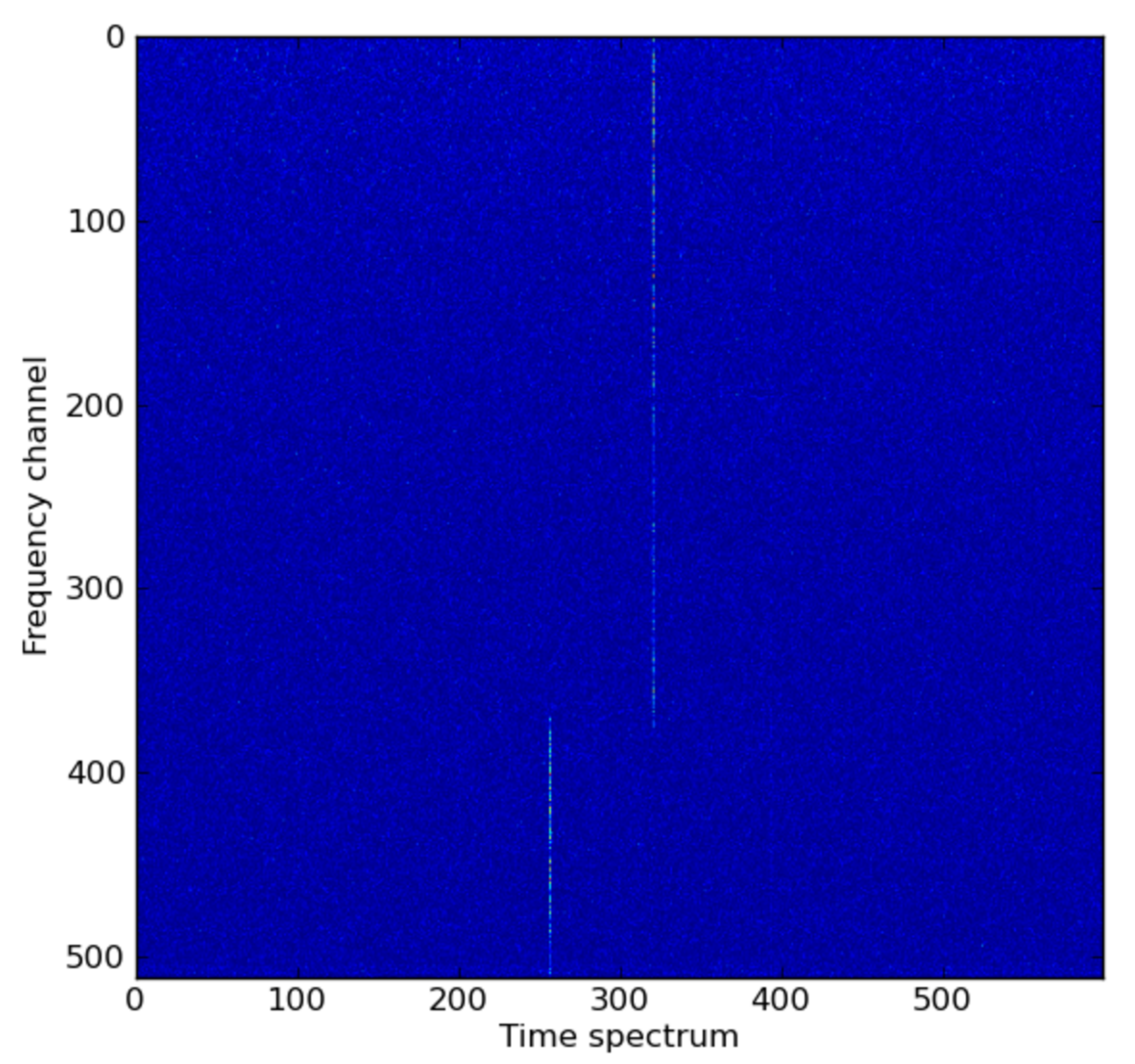}}
  \hspace{8mm}
  \subfloat[Thresholded broadband RFI spike]{\includegraphics[width=200pt]{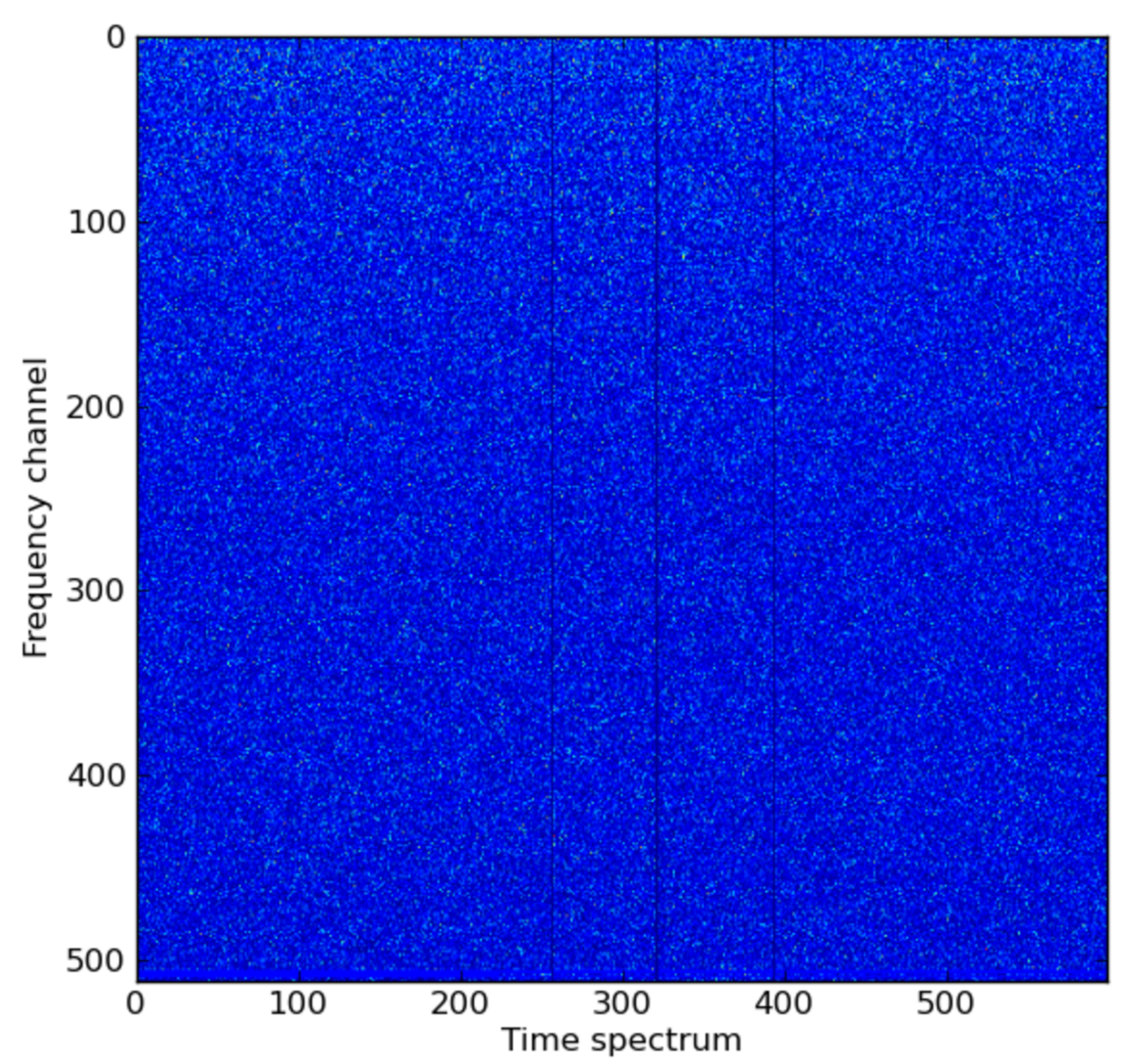}}
  \hspace{8mm}
  \subfloat[Pulse with narrowband RFI]{\includegraphics[width=200pt]{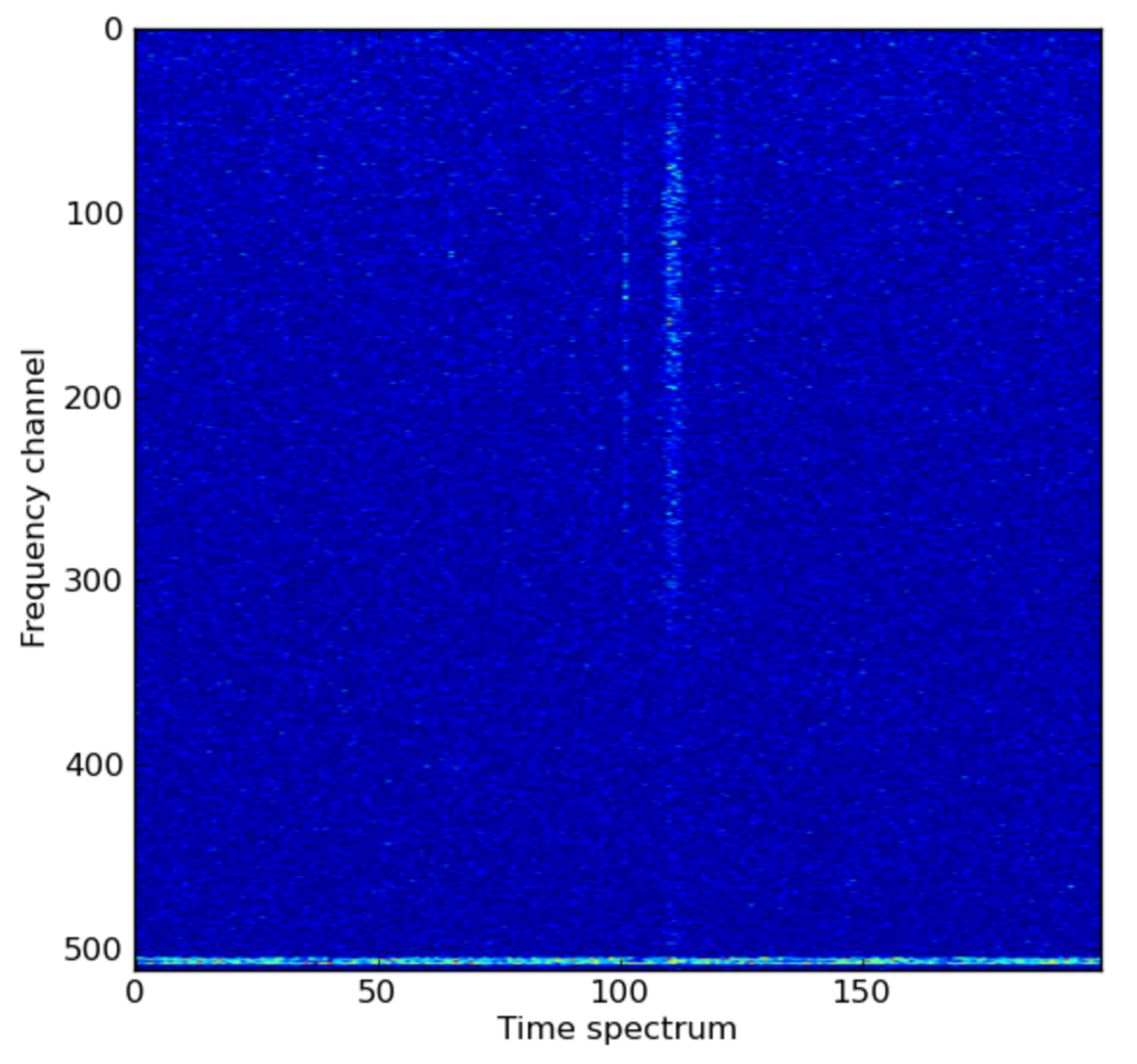}}
  \hspace{8mm}
  \subfloat[Narrowband RFI thresholded]{\includegraphics[width=200pt]{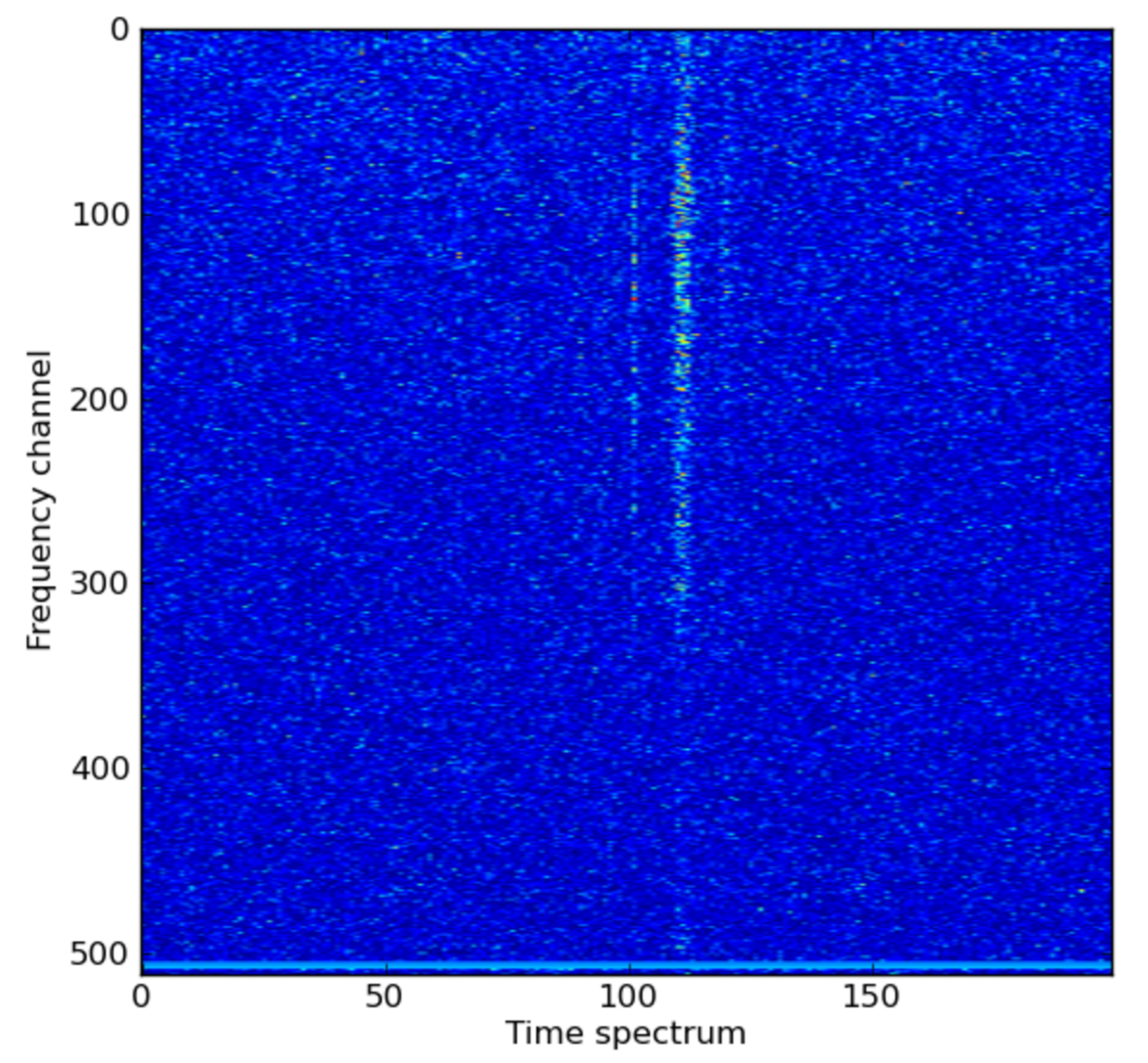}}
  \caption{RFI events during a test observation using a 10MHz bandwidth between 413 and 403 MHz. Plot (a)
	  shows three broadband spikes affecting spectra at 260, 330 and 395, while (c) shows narrowband 
	  RFI affecting frequncy channels 504 to 508. Plots (c,d) also contain a 
           transient signal originating from pulsar PSR B0329+56 which was being observed during the event. 
	   For both cases all thresholding stages were enabled and had the same threshold parameters. 
	   The power intensity values are arbitrary.}
  \label{medicinaRFIFigure}
\end{figure}

Terrestrial RFI, and methods to try and mitigate or correct for it, is one of the major problems in transient surveys,
especially for radio telescopes which are close to urban centers, such as Medicina. Several methods have been
investigated to cater for this (for example, \cite{hogden12}), the major challenge being to try and remove
as much RFI as possible without affecting any true astrophysical dispersed pulses which might be present in the data. 
We implement a simple RFI-thresholding process which limits the amount of strong, bursty RFI,
whose effectiveness can be tweaked by applying different thresholding factors. Any undetected RFI will result 
in incorrect detections after dedispersion. Although these are never welcome, it would be more advantageous to allow 
some low-power RFI to seep through rather than increasing the probability of thresholding astrophysical events. The detection 
and classification stage should then be able to discern between real and terrestrial signals. Not performing any RFI mitigation would 
result in a large number of incorrect detections as well as a higher variance in the data, which lowers the probability of 
detecting weak astrophysical signals. The thresholding process consists of the three stages described below, and figure 
\ref{medicinaRFIFigure} depicts some examples of RFI events which occurred during test observations 
and the resultant waterfall plots after RFI rejection.

\begin{figure}[ht]
\begin{center}
\includegraphics[width=350pt]{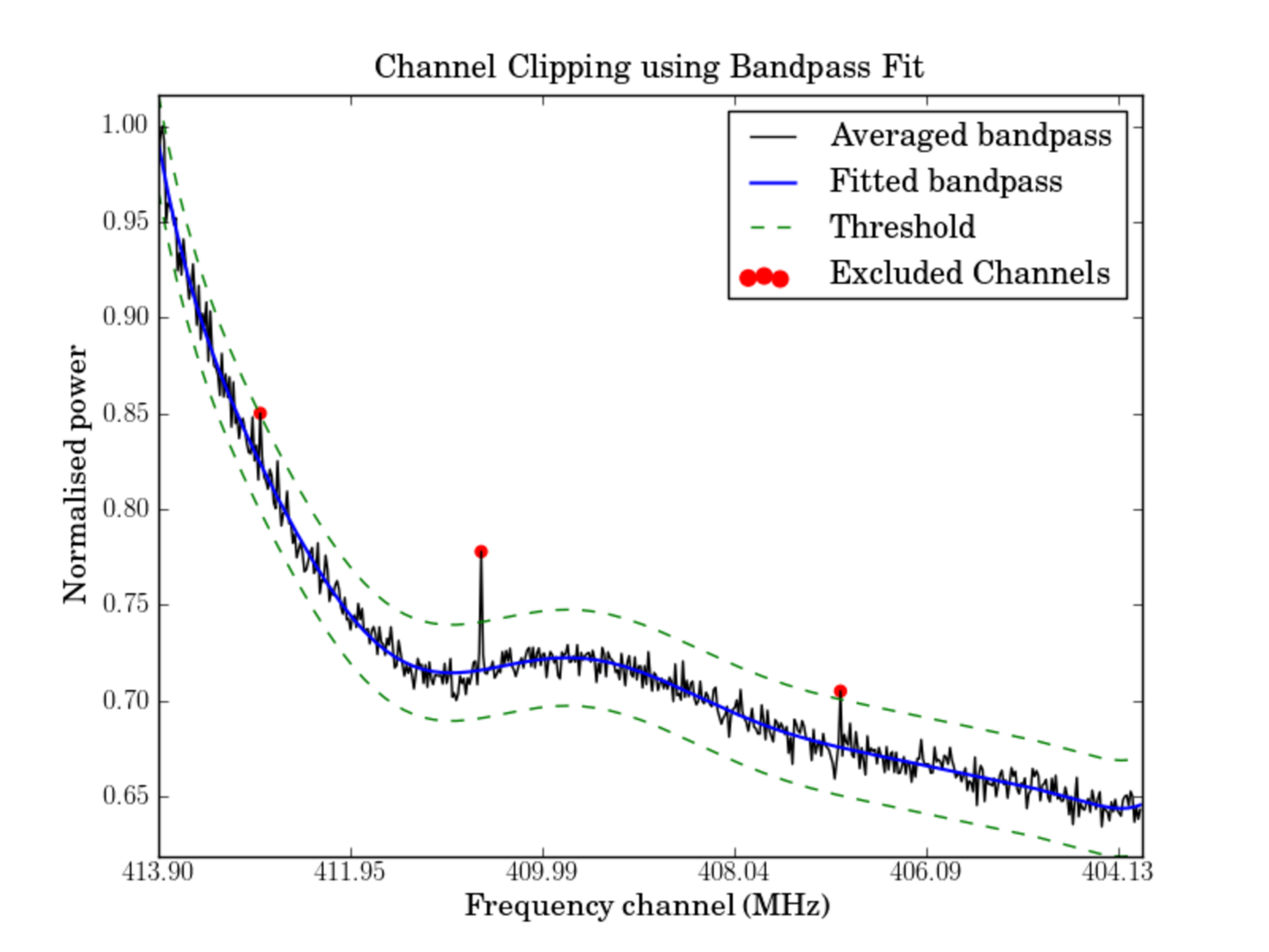}
\end{center}
\caption{The averaged bandpass (black) is fitted with a $p^{th}$ order polynomial (blue) and the RMSE of the two
	 is used to generate channel thresholds (green). This threshold is then applied to subsets of spectra
         of size $W_c$ and any chunk exceeding it will be replaced with the channel's fitted bandpass value, as is the
         case for the channels marked in red in this plot.}
\label{bandpassFigure}
\end{figure}

\begin{description}
 \item{\bf Bandpass Correction:} A $p^{th}$-order polynomial is fitted to an averaged bandpass in each iteration,
           providing a smoothed, approximate description of the telescope's response to the frequency
           band being processed. Accumulation and averaging is performed on the GPU, mainly consisting of a
	   reduction sum across the frequency channels, resulting in $N_{chans}$ values which are copied back to CPU 
	   memory. Channels containing high power RFI will distort the bandpass fit, so these channels are masked
           beforehand by replacing their values with the mean of the neighboring channels, thus interpolating the bandpass.
	   The resulting bandpass is then fitted using least-squares.
	   The root mean square error (RMSE) between the fitted bandpass and the averaged, interpolated bandpass is then
	   computed and used as a thresholding factor for the channel thresholding stage, whilst the mean and standard
	   deviation are used for the spectrum thresholding stage. The fitted bandpass is then subtracted from all the 
	   spectra on the GPU, generating corrected spectra, having
	   the effect of equalizing the telescope's response. This also has the effect of moving
	   the mean to, or near, 0.
 \item{\bf Channel Thresholding:} Each frequency channel should have a uniform power level across short time spans, and if 
           it experiences a sudden change it can generally be attributed to narrow-band RFI (or alternatively, a very
           strong astrophysical burst). First a non-overlapping sliding window of width $W_c$ is applied to each
	   frequency channel, thereby partitioning them into chunks of width $W_c$. The mean of each chunk is calculated,
	   and if this exceeds a threshold then it, together with its immediate neighbors, is flagged as RFI. After
	   the flagging stage is complete the values of any flagged chunk is replaced with the frequency channel's 
	   fitted bandpass value. This is preferred to 
	   setting these values to 0 since the effect on the statistical properties of the data is minimized. $W_c$
	   depends on the RFI environment, the observing parameters and the width of astrophysical 
	   transient being searched for.
 \item{\bf Spectrum Thresholding:} In this stage, the mean of each full-band spectrum is calculated and compared to 
           the spectrum threshold. If the mean exceeds this threshold then the entire spectrum is substituted with 
           the fitted bandpass. Dispersed pulses should not be affected by this process, unless the threshold is
	   exceptionally low, since their power is distributed across multiple spectra depending on the pulses' DM value.
	   The threshold is empirically set such that time spectra are rarely affected by this stage, although this will
	   depend in the RFI environment.
\end{description}

\subsubsection{Dedisperison}

By far the most time-consuming processing stage is dedispersion, which corrects for the frequency-dependent
dispersion experienced by electromagnetic radiation emitted from a source as it travels through the 
Inter-Stellar Medium (ISM). This dispersion obeys the cold plasma dispersion law, where the time-delay 
between two frequencies $f_1$ and $f_2$ is given by the quadratic relation \cite{lorimer05}:
\begin{equation}
\label{dispRelationshipEquation}
 \Delta t \simeq 4.15 \times 10^6 \mbox{ ms} \times \left( f^{-2}_1 - f^{-2}_2 \right) \times \mbox{DM}
\end{equation}
where DM is the dispersion measure in pc cm$^{-3}$ and $f_1$ and $f_2$ are in MHz. Astrophysical objects have 
a signature DM value associated with them, which is related to the direction-dependent distance from the observer.
Brute-force dedispersion is an $\mathcal{O}\left(N_{chans} \cdot N_{samp}\right)$ operation, where $N_{chans}$ 
is the number of frequency channels and $N_{samp}$ is the number of input spectra, which for real-time streaming applications
can be seen as infinite. During a blind transient survey this operation has to be performed over a large number of DM 
values, depending on the observation and survey parameters (see \cite[Section 2]{magro11}, for an explanation of how 
to generate search parameters). We use the dedispersion kernel developed in \cite{armour12} where each GPU thread 
block is responsible for processing an area of $(DM,t)$ space, maximizing data locality in the fastest memory 
locations.

\subsubsection{Post-Processing}

The dedispersed time series are then passed through a post-processing stage, which smoothens and normalizes
the data. This is mostly done using two techniques:

\begin{description}
 \item{\bf Median Filtering: } Noisy, isolated outliers in the time series can be suppressed by replacing the 
	    value $y_m$ with one which is calculated using neighboring points:
	    \begin{equation}
	      \label{medianFilterEquation}
               y_m = operation\{x_i, i \in w\}
	    \end{equation}
	    where $w$ represents a neighborhood centered around the location $m$, and $operation$ is the averaging 
            function used. We chose to use the median of the neighborhood points, since it better preserves the 
            edges of the signal than the mean. We have implemented a windowed-median filter on the GPU, where thread
            blocks are split across a 2D grid with each row processing one dedispersed time series, for a single
            DM value, partitioned along the columns of the grid. Each thread block loads a subset of the series to shared
            memory, including some overlapping values at the edges, and each thread computes the median of its associated
            data point neighborhood and stores this value back to global memory. 
 \item{\bf Detrending and Normalization:} The mean power of the incoming data stream can change gradually in time, the
            rate of which depends on the cause of this change, such as steady temperature changes in telescope electronics, 
	    sky temperature variations, or an astrophysical radio source moving toward, or away from, the beam center. This
            effect can be alleviated by subtracting a best-fit line to the dedispersed time series, which effectively 
	    centers the series to a mean of 0. The detrending process is also performed on the GPU, where a thread block
            is associated with one time series and the best-fit line is computed using linear regression. The kernel requires
            two passes of the data, one to calculate the regression parameters and one to subtract the trend-line from the
            series. During this pass, the standard deviation is also computed, and a third pass is performed to normalise
            the data.
\end{description}

After this stage, the post-processed dedispersed time series are copied back to CPU memory and forwarded to the event 
detection stage, which is started as soon as all the beams have been processed (and a new input data buffer is available
for processing).

\subsection{Event Detection}

During the first stage of event detection, the dedispersed time series are thresholded using a suitable threshold value ($n\sigma$).
This value should be low enough to allow low SNR pulses to pass through, even at the cost of incorrect RFI detections, which will
be filtered in the clustering stage. A list of detections is generated for each beam, containing $\left(time,\;DM,\;intensity\right)$
triplets. Astrophysical transients, as well as RFI signals, will result in a number of entries in this list which should be grouped
together and treated as a single candidate. This is the main function of the clustering stage.

We start by applying a density-based clustering technique, DBSCAN \cite{ester96}, to group neighboring data points together. Its 
definition of a cluster is based on the underlying estimated density distribution of the dataset. The shape of the clusters is 
determined by the choice of the distance function for two points. The {\it Eps-neighborhood} of a point $p$, denoted by $N_{Eps}(p)$
is defined by $N_{Eps}(p)=\{q\in D | dist(p,q) \leq Eps\}$ where $Eps$ is an argument defining the neighborhood extent of a point, $D$
is the set of points and $dist(p,q)$ is the distance function for points $p$ and $q$. DBSCAN distinguishes between two types of cluster
points, points inside the cluster ({\it core points}) and points on the border of the cluster ({\it border points}) which generally have 
less points in their neighborhood. A point $p$ is {\it directly density-reachable} from $q$ if $p 
\in N_{Eps}(q)$ and $|N_{Eps}(q)| \ge MinPts$. A point $p$ is {\it density reachable} from a point $q$ if there is a chain of points 
$p_1,...,p_n$, $p_1=q$, $p_n=p$ such that $p_{i+1}$ is directly density-reachable from $p_i$. Border points within the same cluster 
might not be density-reachable from each other, but they are {\it density-connected} if there is a point $o$ such that both of them 
are density-reachable from $o$. Following these definitions, a {\it cluster} is defined as a non-empty subset of $D$ satisfying the 
following two conditions:

\begin{unnumlist}
 \item  {\it Maximality}: $\forall p,q : p \in C \wedge (q\mbox{ is density-reachable from } q) \Rightarrow q \in C $   
 \item {\it Connectivity}: $\forall p,q \in C : p\mbox{ is density-connected to } q$
\end{unnumlist}

Any points which do not belong to a cluster are regarded as noise: $N = \{p \in D | \forall i:p \notin C_i\}$, where $N$ is the set of 
noise points, $C_1,...,C_k$ are the clusters in $D$ and $i=1,...,k$ with $k$ being the total number of clusters. This technique has several 
advantages  which makes it a suitable candidate for clustering detections:
\begin{arabiclist}
 \item it does not require a seed to specify the number of clusters in the dataset, which is a desirable property since the number of 
       events occurring within a time-frame, be they astrophysical transients or RFI, is unknown unless a cluster approximation step is 
       performed beforehand
 \item it has a notion of noise
 \item it is also capable of separating overlapping clusters having different density distributions, such as when an RFI signal 
       overlaps a transient event, if the input arguments are sensitive enough
\end{arabiclist}

The main drawback of this technique is its runtime complexity, dominated by the neighborhood calculation for every point, which is of order 
$\mathcal{O}(N^2)$ unless an indexing structure is used or a distance matrix is computed beforehand, however this needs $\mathcal{O}(N^2)$
memory, which is unfeasible for large datasets. To counter this we implement a faster, 
albeit less accurate, version of the algorithm, FDBSCAN \cite{zhou00}, which only uses a small a subset of representative points in a 
core point's neighborhood as seeds for cluster expansion, reducing the number of region query calls. The representative points are chosen 
to be at the border of a core point's neighborhood, two for each dimension, one for each direction when placing the core point at the origin.
Due to this approximation, some points might be lost, and in some cases clusters might be split apart, however the probability of this 
happening is very low (see \cite{zhou00}). 

The distance function used in our implementation assigns a different value to each dimension ($T_t$, $SNR_t$, $DM_t$). Detections after 
dedispersion will have a specific shape along the time dimension due to incorrect dedispersion, with higher DM values detecting events 
before lower ones. The width of a cluster will also reflect the true pulse width, being narrower near the true DM.
The value of $T_t$ should depend on the lowest pulse width being searched for. A higher $T_t$ might result in pulses close in time to be 
fused together into one cluster. The value of $SNR_t$ and $DM_t$ should be large enough to allow clusters to encompass the entire range,
whilst allowing DBSCAN to distinguish between core and border points.

The output of the clustering stage is a list of detected clusters. The main challenge is then to discern between RFI-induced clusters 
and transient candidates. In the case of transients the highest SNR detections will be centered around the pulse's true DM, dimishing
in power when moving away from the value until the threshold level is reached due to dedispersion at an incorrect DM value. Following 
\cite{cordes03}, assuming a rectangular bandpass function, which should resemble the telescope's bandpass shape after bandpass
correction, a Gaussian-shaped pulse with FWHM width $W$ in milliseconds, the ratio of measured peak flux density $S(\delta DM)$
to true peak flux $S$ for a DM error $\delta$DM is

\begin{equation}
 \frac{S(\delta DM)}{S}=\frac{\sqrt{\pi}}{2}\zeta^{-1}erf\zeta
\label{incorrectDedispersion}
\end{equation}
where
\begin{equation}
 \zeta = 6.91\times10^{-3}\delta DM \frac{v_{MHz}}{W_{ms}v^3_{GHz}}
\end{equation}

Comparing this model with a cluster's DM-SNR signature provides us with a classification mechanism. Our implementation
performs the following processes for each detected cluster:

\begin{enumerate}
 \item Generate its DM-SNR signature by collapsing the time dimension
 \item Smoothen this signature by running a $k$-element moving average
 \item Find the DM value containing the highest number of detections and its maximum SNR. If this DM value is less than 1.0 then it is assumed 
       that it was caused by broadband RFI and the cluster is discarded
 \item Approximate the pulse's FWHM by computing the difference between the pulse's start and end time at the maximum SNR value.
 \item Normalise SNR-DM signature
 \item Compue the analytical curve for incorrect dedispersion using equation \ref{incorrectDedispersion}
 \item Calculate the RMSE between the modelled curve and pulse's DM-SNR signature
 \item If the MSE exceeds a preset threshold, then the cluster is discarded (RFI), otherwise classify as a potential candidate. This 
       threshold can be set empirically through test observations or by using simulated data.
\end{enumerate}

This procedure is most effective for detecting relatively strong pulses and differentiating them from RFI. The classification of clusters 
having a small number of detections can be incorrect if the width of the pulse cannot be determined. To increase the likelihood
of detecting lower-SNR pulses, the detection threshold during the first stage of event detection should be lower than typically 
used for similar searches in order to allow more pulse detections to be clustered together. This will result in a higher number of 
background noise detections, however these will be filtered out by DBSCAN.

Figure \ref{classificationFigure} depicts the output of this process when applied to detections during a test observation with
BEST-2 (see section \ref{deployment}). Selected candidates are flagged and the data buffer in which they were found is persisted to 
disk together with cluster information.

\subsection{Data Writer}

\begin{figure}
  \centering 
  \subfloat[]{\includegraphics[width=235pt]{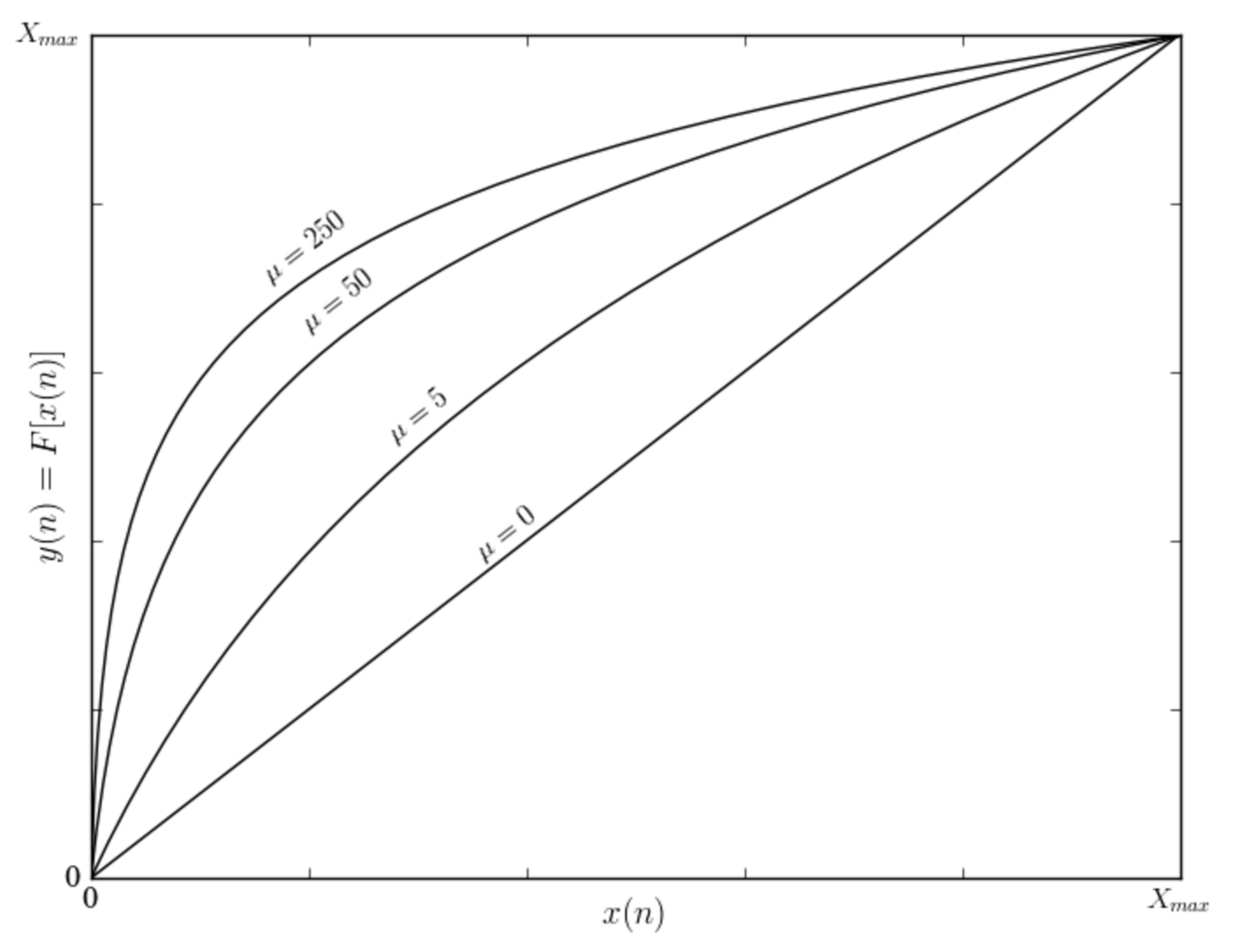}}
  \hspace{8mm}
  \subfloat[]{\includegraphics[width=235pt]{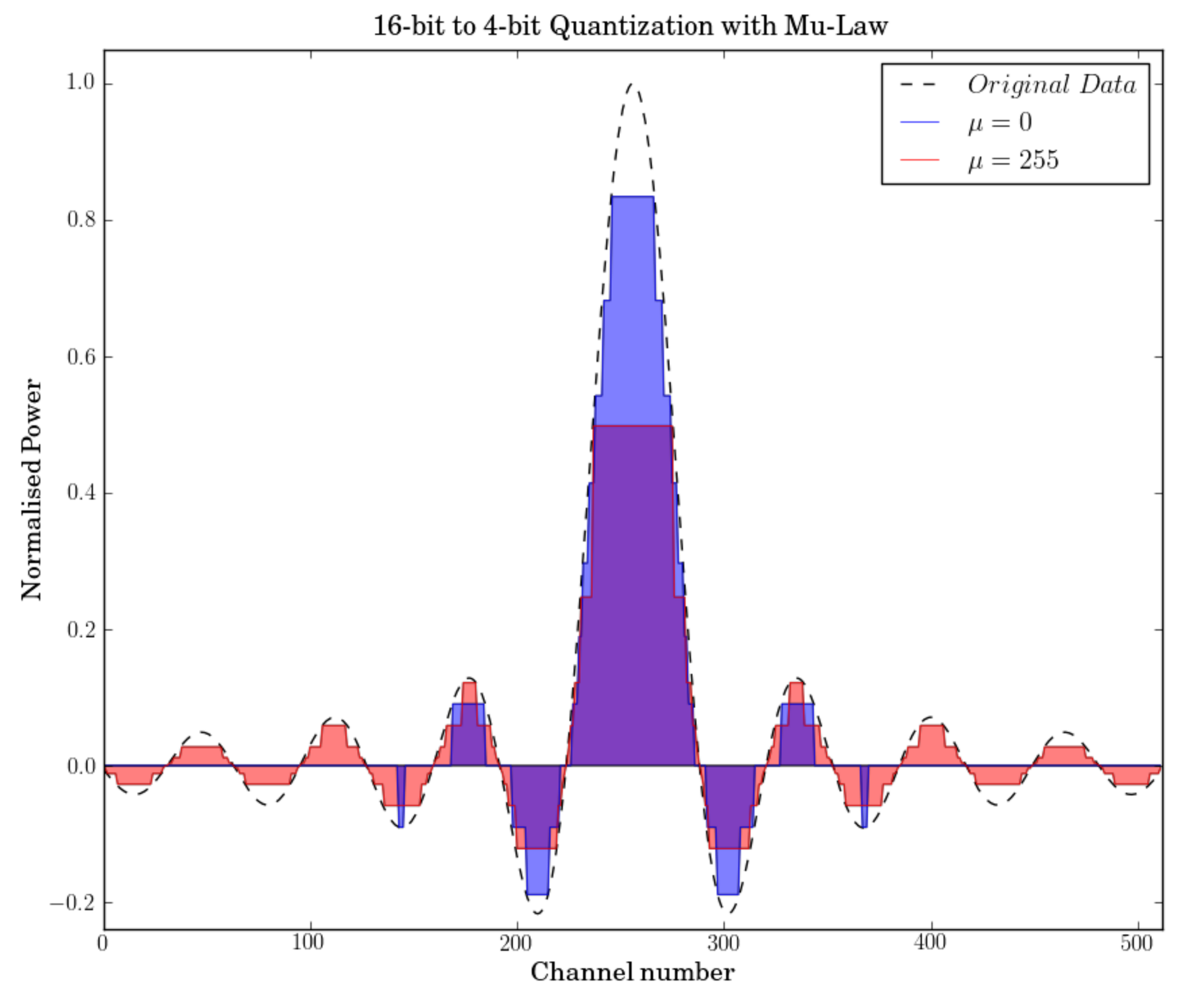}}
  \caption{Plot (a) depicts the mapping between input and output values for a given range when applying the $\mu$-law algorithm,
	  while plot (b) shows the effect this has when quantising signed 16-bit input values down to signed 4 bits. When the compression
	  factor is high most of the detail in the trailing curve is retained, with the expence of losing information at higher values.
	  Applying this to channelised raw voltages would have the effect of allocating more bits to the region around the mean and clipping 
	  high-valued outliers. }
  \label{muLawFigure}
\end{figure}

The pipeline can either write the entire incoming data stream to disk or dump data buffers containing interesting detections
when triggered by the event detection stage, for future off-line processing. The data rates being processed can go
up to 650 MB/s, which is much faster than what conventional hard drives can process, so the data has to be quantized
first. This data can be in two formats: 
complex channelized time series, in which case 16-bit complex values are quantized to 4-bits with two complex components
packed into one byte, or channelized power series, where each 32-bit single-precision floating point value is quantized
to 8 bits.

In both cases, although the dynamic range can potentially span the entire bit-range, most of the values will typically be 
distributed across a small range around a mean level (in the simplest case, complex channelized series will be normally
distributed, while channelized power series will follow a half-normal distribution). Using a linear quantizer would
result in a loss in sensitivity within this region, whilst clipping the range would reduce the SNR of any potential
pulses in the data. For this reason, we use a logarithmic quantizer, specifically a $\mu$-law quantizer, adapted
from the G.711.0 standard \cite{ITUT}:
\begin{equation}
 y = X_{max}\frac{log\left[1+\mu\frac{|x|}{X_{max}}\right]}{log\left[1+\mu\right]}sign\left[x\right]
\end{equation}
where $x$ represents points in the data series, $X_{max}$ is the maximum value in this series, $\mu$ is the compression
factor and $y$ is the quantized data series. A higher compression factor will result in more output bits being allocated 
to the high data point concentration range of the input distribution, as depicted in figure \ref{muLawFigure}. The data 
buffer is first encoded using this mapping and then the output values are quantized and dumped to disk. This is performed 
on the CPU, so as not to interfere with the main processing pipeline and induce delays. A precomputed lookup table 
for $log$ values is used to speed-up processing. The data series is split across multiple OpenMP threads, whose processing 
is interleaved with file I/O calls, in order to overlap CPU-processing and I/O.

\section{Performance Benchmarks}
\label{benchmarks}

\begin{figure}
  \centering 
  \subfloat[Clustering scaling benchmark]{\includegraphics[width=235pt]{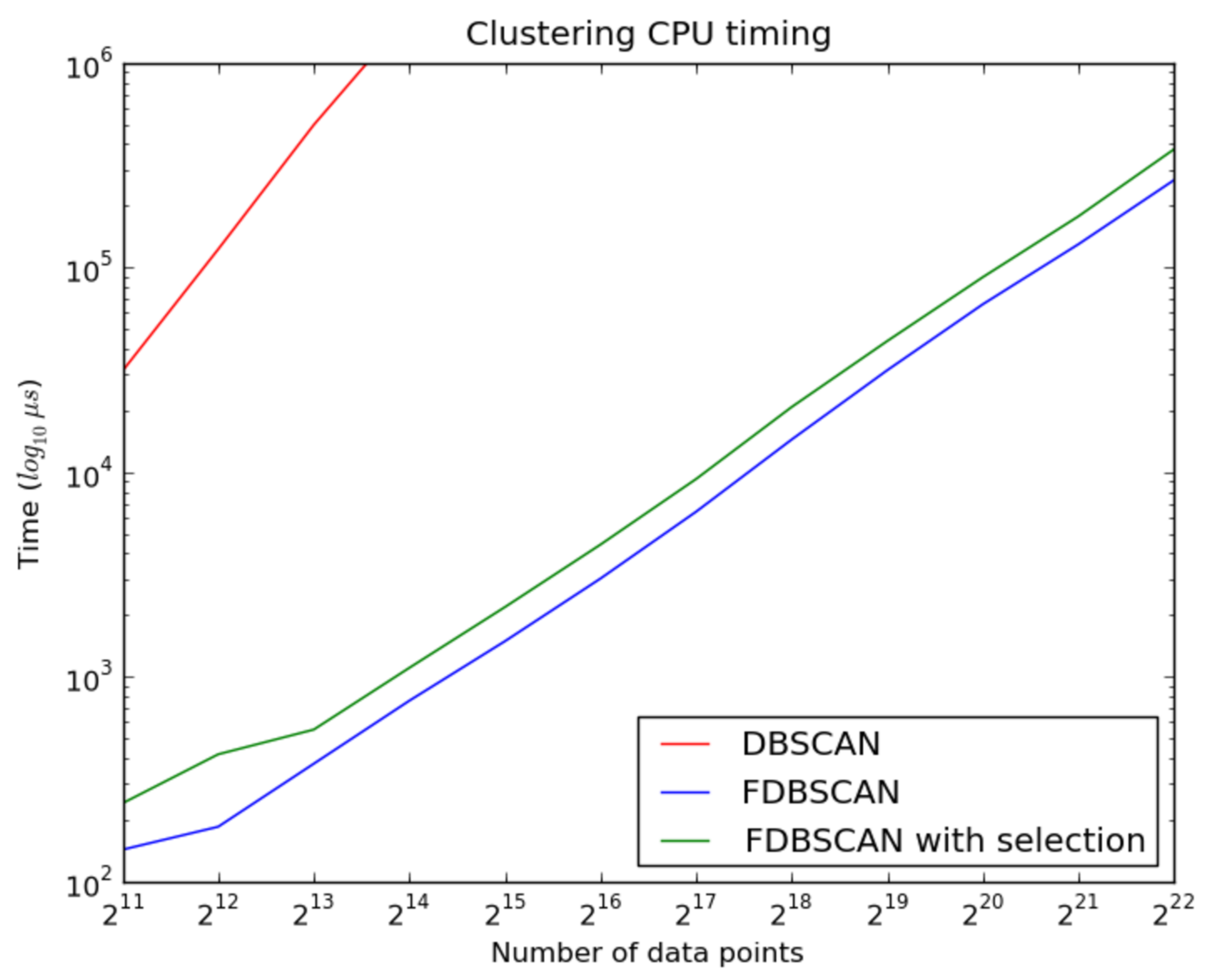}}
  \hspace{8mm}
  \subfloat[Quantization scaling benchmark]{\includegraphics[width=235pt]{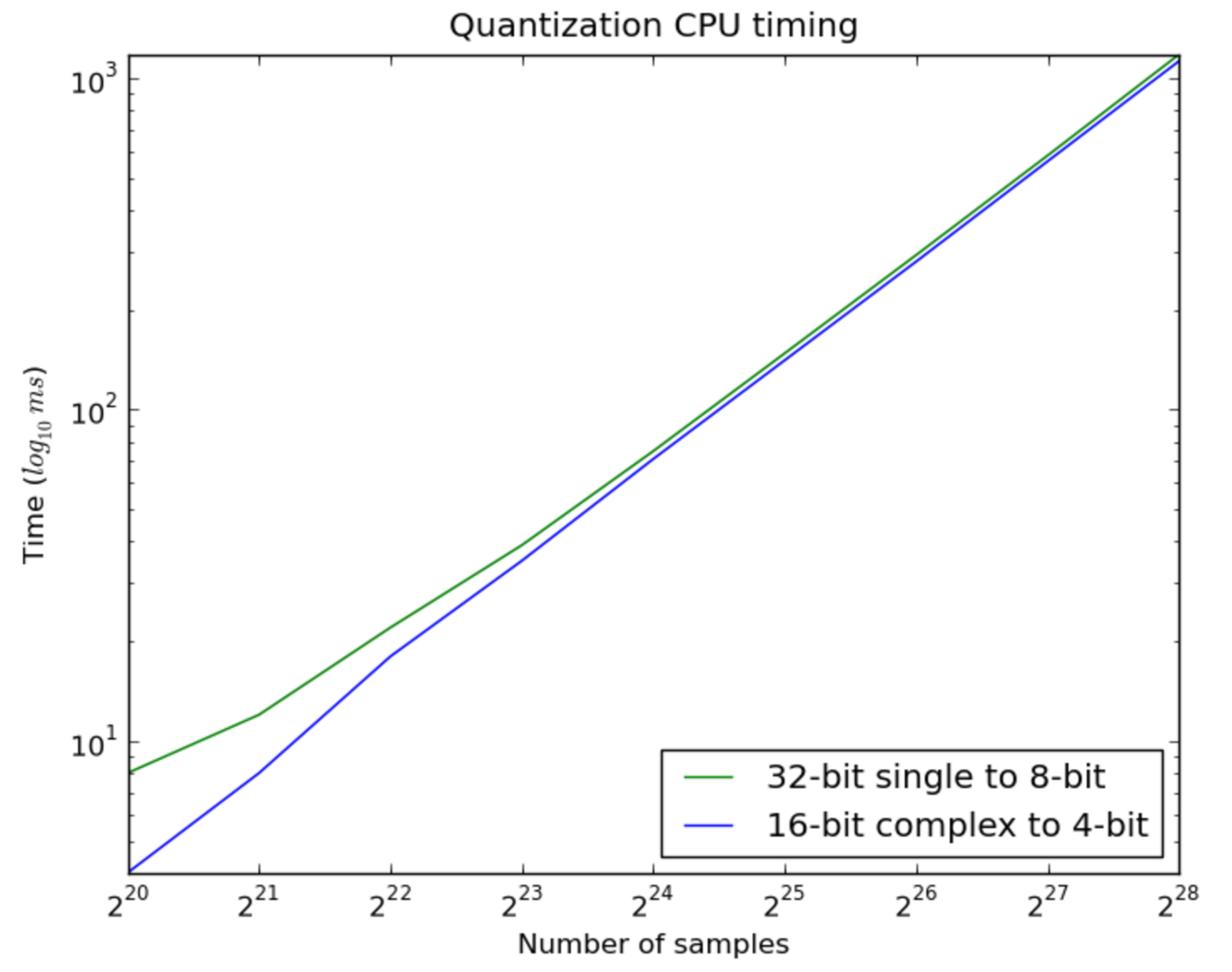}}
  \caption{Scaling benchmarks for clustering and quantization on the CPU. Plot (a) shows the significant difference in runtime between
	   DBSCAN and FDBCAN, with candidate selection only increasing this by a small factor. Initial observations and empirical testing
	   shows that the number of data points to cluster rarely exceeds $2^{21}$, making the implementation about an order
	   of magnitude faster than real-time. Plot (b) depicts the linear scaling for quantization when converting 32-bit single precision
	   values to 8-bit and two 16-bit components to two 4-bit values packed into one byte. When processing 8 full bandwidth beams, the number
	   of samples to process is approximately $2^{27}s^{-1}$, which can be performed in real-time when excluding the time required to 
	   dump the encoded data to disk.}
  \label{cpuTimingFigure}
\end{figure}

\begin{table}[]
  \centering
  \begin{tabular}{ | p{3.2cm} | r | p{3.2cm} | r | }
    \hline
    \multicolumn{4}{|c|}{Copy to GPU: 295.30} \\
    \hline
    \hline
    \multicolumn{2}{|c}{GPU} \vline & \multicolumn{2}{c|}{CPU} \\ 
    \hline
    Bandpass Fitting & 36.79 ms   & Thresholding     & 690.96 ms  \\
    RFI Filtering    & 74.97 ms   & Clustering       & 1148 ms \\
    Dedispersion     & 3863 ms    & Classification   & 168.64 ms  \\
    Median Filtering & 135.11 ms  & Clusters to file & 728.73 ms  \\
    Detrending       & 59.06 ms   & Quantization\footnotesize{$^*$}  & 3532 ms     \\
    \hline
    \hline
    \multicolumn{4}{|c|}{Copy from GPU: 155.97} \\
    \hline
    \hline
    \multicolumn{4}{|c|}{Total iteration time: 4619.72ms} \\
    \hline
  \end{tabular}
  \caption{GPU and CPU timings for one pipeline iteration when processing 8 full bandwidth (20MHz) beams split between 2 GTX 660Ti. 
	   GPU timings are for a single GTX 660 Ti processing 4 beams while CPU timings are for all the beams when processed using 
	   a single output thread and quantization thread. The data buffer contains approximately 5s of channelised power data with 
	   a simulated 500ms periodic pulse with 1\% duty cycle and an SNR of 3, resulting in about 18 clusters with a total of $2^{16}$ 
	   points. The maximum possible number of DM values which can keep up with real time is 640, and in this case the maximum 
	   DM is 102.3 $pc\ cm^{-3}$. 
	   \newline		
	  \footnotesize{* Quantisation is performed on a different CPU thread from the event detections stages}}
  \label{timingTable}
\end{table}

\begin{figure}[ht]
\begin{center}
\includegraphics[width=390pt]{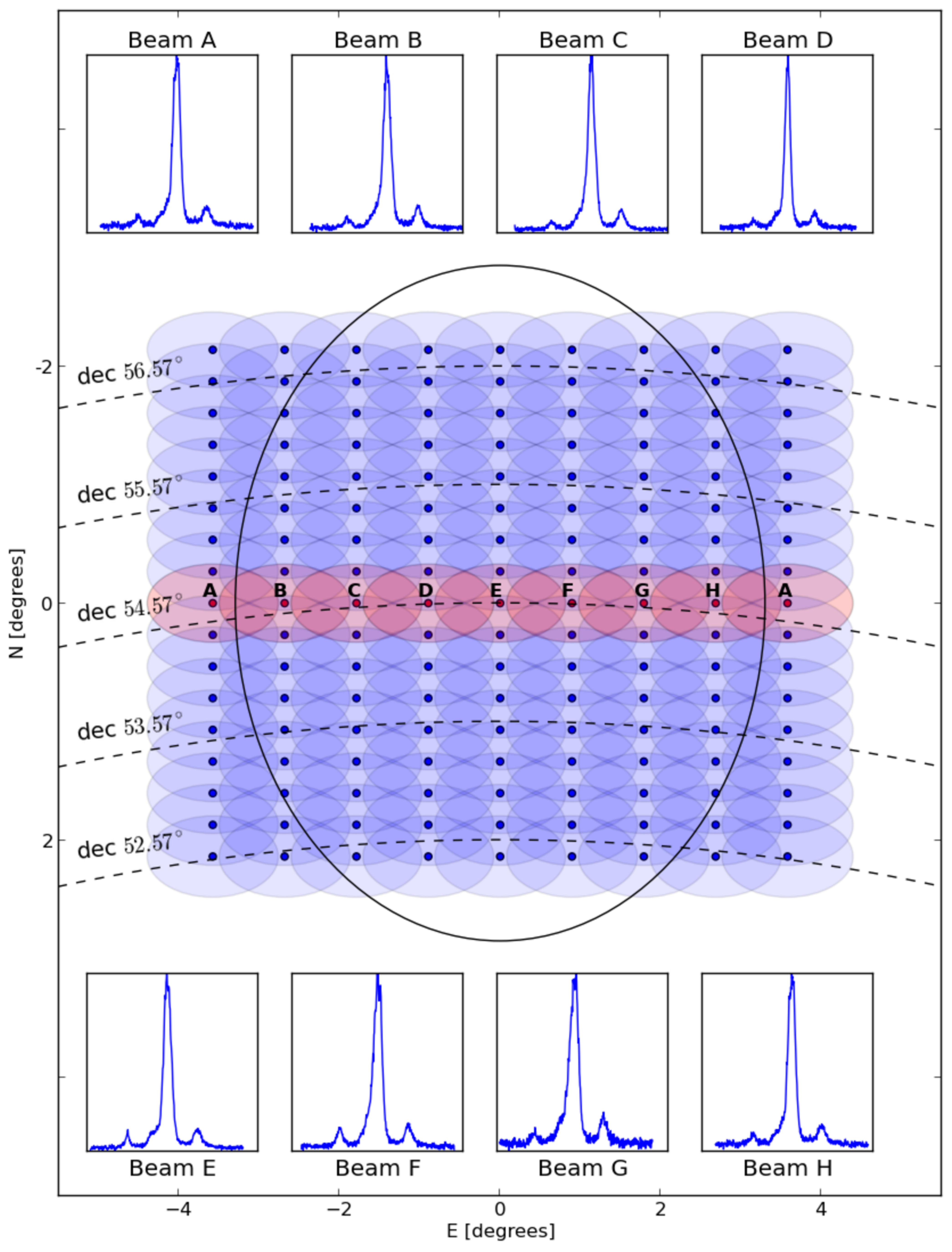}
\end{center}
\caption{The digital FFT beamformer generates a 2D grid of synthesized beams, out of which eight can
         be selected for output. The centers of each beam, together with their FWHM, are shown in this plot.
	 Beams get wider further from the zenith, which for BEST-II is at 44.52$^\circ$. During test observations
	 the central E-W row of beams is chosen, so that a transient source will transit through as many beams
	 as possible. The figure above shows the folded pulse profiles from an observation of PSR B0329+54, which
	 transits all the beams during a 1800s observation, its trajectory depicted by the dashed line at a DEC
         of 54.57$^\circ$. The channelized complex data from each beam was quantized and persisted to disk, after
	 which the integrated pulse profile for each beam were generated using 50 profiles. The differences between the profiles
	 can be attributed to some RFI events which occurred during the observation.}
\label{beamConfigurationFigure}
\end{figure}

This pipeline can be thought of as a soft real-time system, where all the parallel stages should keep up with the incoming
data stream for maximum quality of service, however if for some reason one of them does not meet a processing deadline the 
system does not become unstable but rather the buffering stage will drop heaps until the pipeline progresses by one
iteration. These processing hiccups might happen when, for example, a very high-power RFI spike induces a large number 
of detections resulting in a large number of data points to cluster. The GPU processing times are fixed regardless of the
quality of the data, so these issues are only applicable to CPU threads. For this reason, a pipeline iteration should be 
limited by the GPU processing time. Figure \ref{cpuTimingFigure} shows the scaling performance for the two most time consuming 
stages on the CPU, clustering and quantization, when being performed by a single host thread. The number of samples which need 
to be quantized for a single pipeline iteration is fixed ($f_{bw}\times N_{beams}\times N_{components} \times t_s $), while the 
number of data points which need to be clustered depends on the number of signals in the data stream, however empirical 
testing and initial observations show that this number rarely exceeds $2^{21}$. Both stages are capable of processing the 
full data stream in real-time for BEST-2 observation parameters using a single CPU thread per stage.

Table \ref{timingTable} lists the timings for all the processing stages for one pipeline iteration when processing 8 full
bandwidth (20 MHz) beams. A simulated data file containing a 500ms periodic pulse with 1\% duty cycle and an SNR of 3
was input to the system, with the data copied to all the beams in $\sim$5s buffers, this length being limited by the amount 
of GPU memory available. The host system on which the benchmark tests were conducted consists of 2 Intel Xeon 
E5-2630 2.3 GHz processors, 2 NVIDIA GTX 660 Ti GPUs with 3GB of GDDR5 RAM, 32 GB DDR3-1600 system RAM and a Fujitsu D3118 
system board. The table is split into two columns which work in parallel, the GPU-based processing stages and the CPU-based
processing stages, with data transfer to and from the two performed at synchronization barriers. It is clear that the 
processing bottleneck is the dedispersion kernel, which takes up 93\% of the GPUs' processing time, while the rest of the 
kernels have a negligible effect on the overall running time of the pipeline. The number of DM values which can be processed
during an observation is limited by this as well, and currently this value is around 640 DMs, which can be doubled when
processing half the bandwidth (see section \ref{deployment}). The DM step and maximum DM do not have a large performance impact,
although a higher DM value would require a larger temporary GPU buffer to store the shifted samples and would limit the number
of time spectra which can be processed in each iteration. For a single beam, the tradeoff between the number of 
spectra and dispersional measure trials which can be processed can be calculated as follows:
\begin{equation}
 N_s = \frac{M \cdot \frac{b}{32} - (\Delta t_{max} \cdot N_c) - \alpha}{N_d + N_c}
\end{equation}
where $N_s$ is the number of time samples, $N_c$ is the number of frequency channels, $N_d$ is the number of DM 
trials, $M$ is the amount of GPU global memory in 32-bit words, $\Delta t_{max}$ is the dispersion shift in 
time samples between the edges of the observing band for the largest DM value, $b$ is the input bitwidth and
$\alpha$ represents other smaller buffers which are required (such as the shift buffer for dedispersion 
and fitted bandpass buffer for bandpass correction).

\begin{figure}
  \centering 
  \subfloat[PSR B0329+54]{\includegraphics[width=230pt]{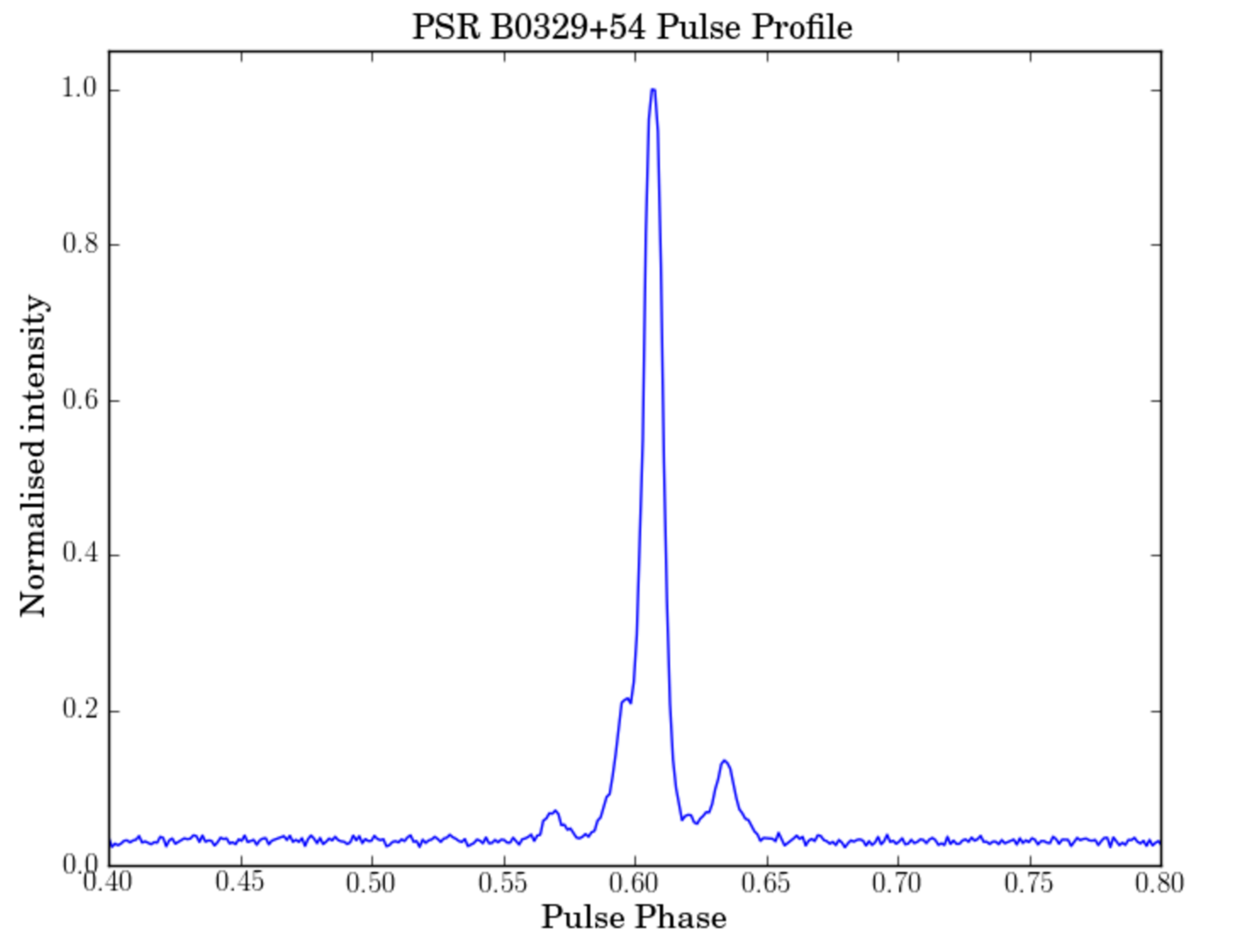}}
  \hspace{8mm}
  \subfloat[PSR B2016+28]{\includegraphics[width=230pt]{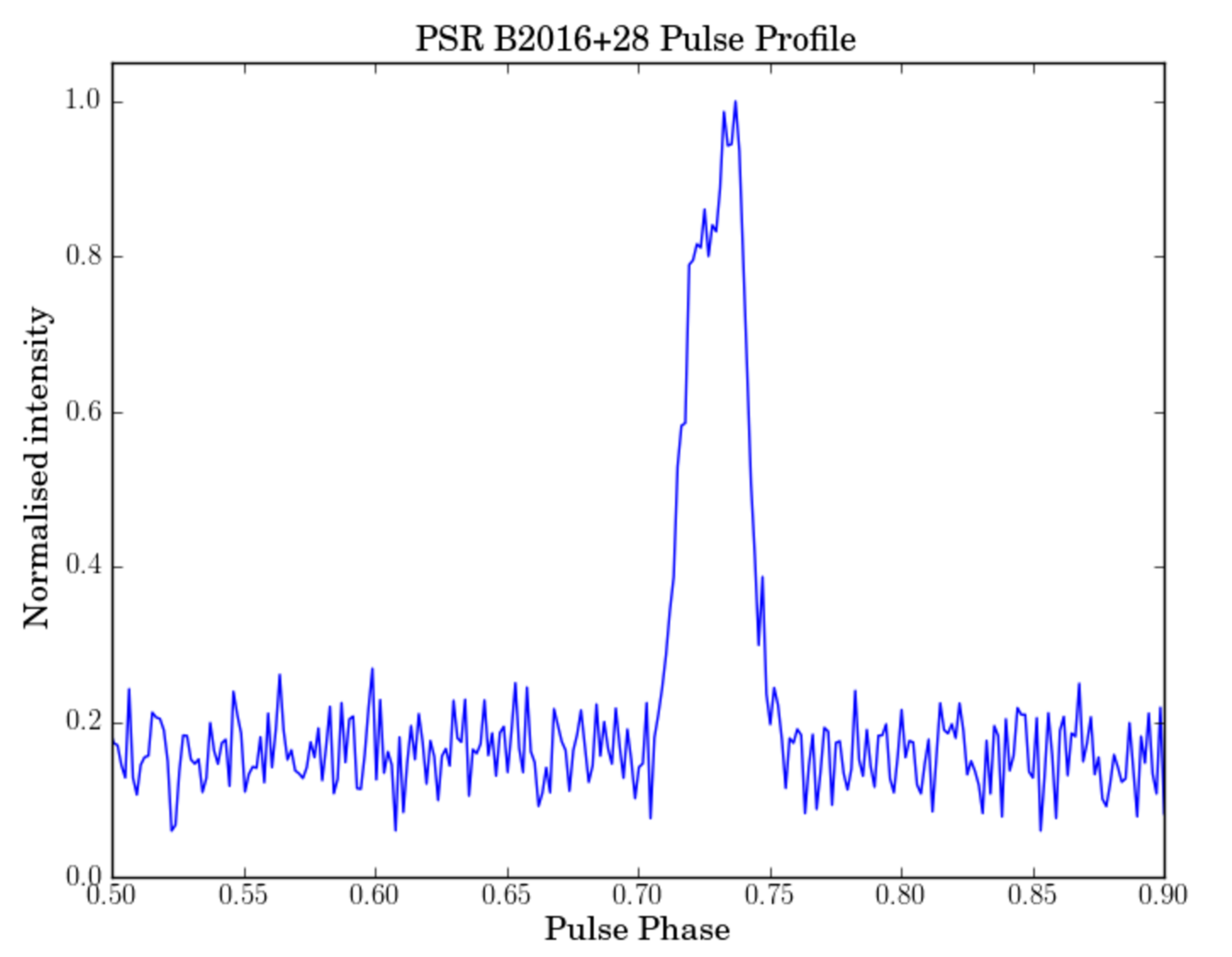}}
  \caption{Integrated pulse profiles for pulsars B0329+54 aand B2016+28 consisting of 200 profiles.}
  \label{profilesFigure}
\end{figure}

\section{Initial Observations}
\label{deployment}

A GPU server (hardware specifications listed in section \ref{benchmarks}) was deployed at the Medicina 
BEST-2 array and connected to the digital backend via a single 10GigE link. Initial test observations indicated 
that the full 20MHz bandwidth contained significant narrowband RFI, with the edges of the bandpass having 
negligible SNR. For this reason, half of the band was discarded and 10 MHz were used for the rest of the observations,
from 413.9 MHz to 403.9 MHz. The bandpass shape is depicted in figure \ref{bandpassFigure}. The beamformer creates 
a 2D grid of beams within the primary beam, out of which eight can be selected for output. These beam were chosen
to create a ``strip'' along the E-W direction (along RA) so that pointing towards a transient source would result
in it transiting across multiple beams, which is useful for testing the digital beamformer, the pipeline setup as 
well as the data transmission between the two. 

\begin{figure}
\begin{center}
\includegraphics[width=440pt]{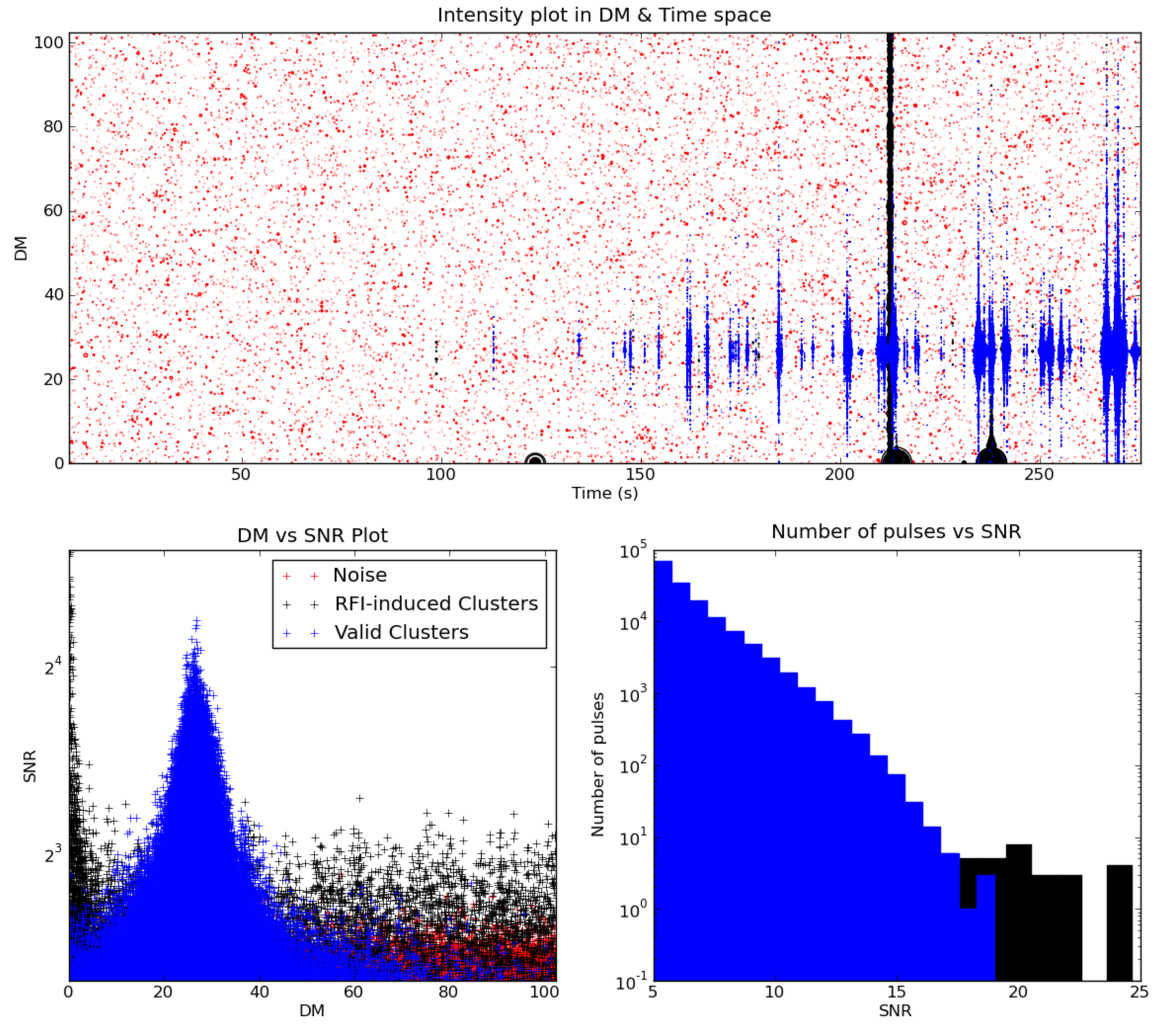}
\end{center}
\caption{A $\sim$280s observation of pulsar PSR B0329+54, which enters the beam at around 100s with the pulse SNR increasing
	 as the pulsar moves towards the center of the beam. This plot shows the output generated by the clustering and 
	 candidate selection stages, partitioning the data into noise (red), RFI-induced clusters (black) and selected
	 candidates (blue). Some of these clusters are also depicted in figure \ref{classificationFigure}.}
\label{b0329Figure}
\end{figure}

Several test observations were performed on known bright pulsars, especially PSR B0329+54, which is the brightest 
transient source that can be observed with BEST-2, located at RA 20:18:03.8333 and DEC +28:39:54.212. The beam 
configuration for these observations is shown in figure \ref{beamConfigurationFigure}, where the central 
row of beams is selected for output (colored in red). The pipeline was used in ``persistence mode'', where 
all the channelized complex-voltages from all the beams are quantized and persisted to disk. The integrated pulse
profiles were then generated for each beam using 50 profiles. The differences between the profiles can be attributed 
to RFI events which occurred during the observation, sometimes resulting in a recalculation of the quantization 
factors (this only happens when an extremely large RFI event occurs, as was the case during this observation).
The integrated pulse profiles for PSR B0329+54 and, through a separate observation, PSR B2016+28 are shown in 
figure \ref{profilesFigure}, each generated by folding 200 profiles offline with raw observation data persisted 
to disk from a single beam.

\begin{figure}
  \centering 
  \subfloat[High SNR pulse from B0329+54]{\includegraphics[width=220pt]{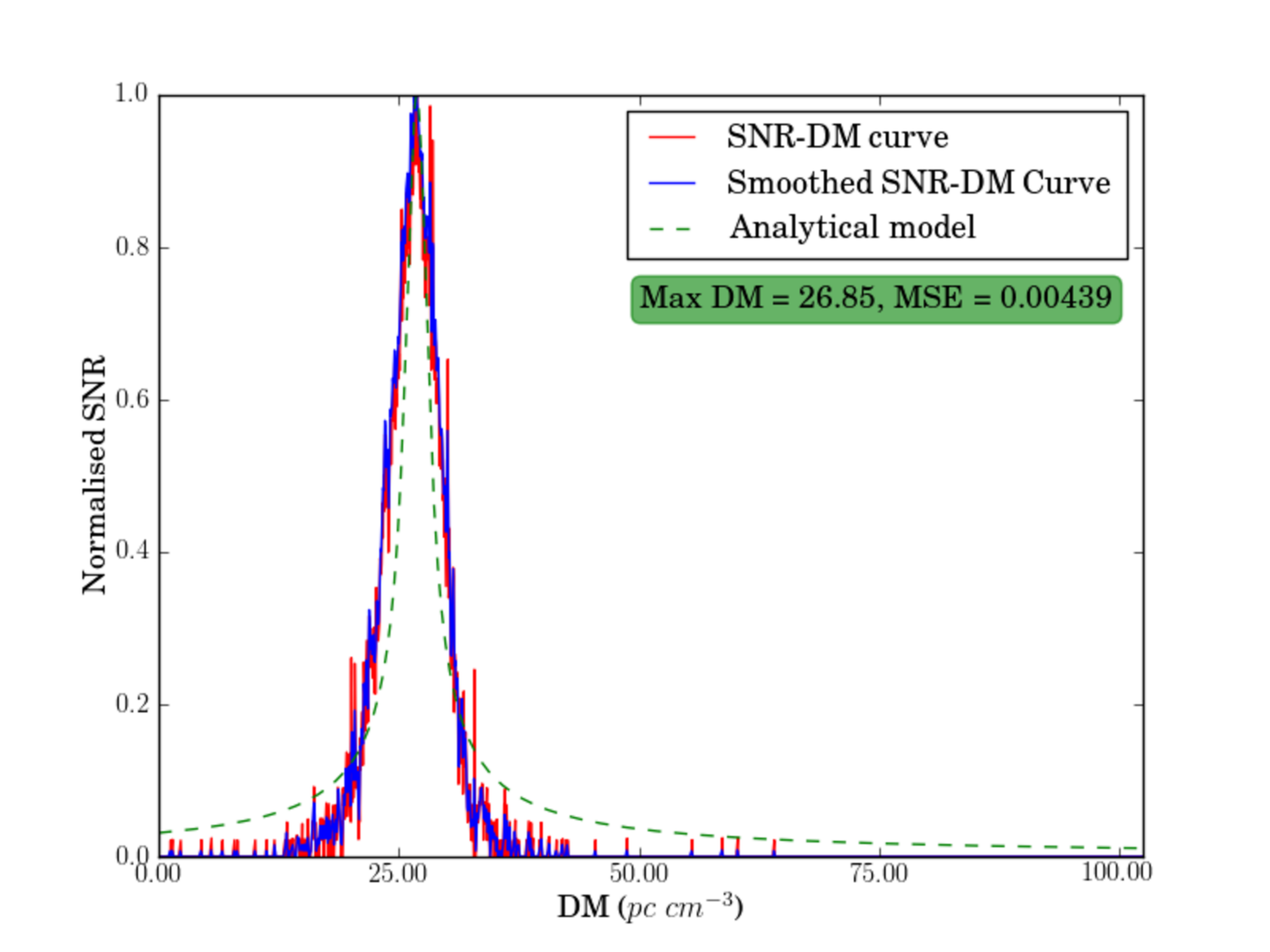}}
  \hspace{8mm}
  \subfloat[Low SNR pulse from B0329+54]{\includegraphics[width=220pt]{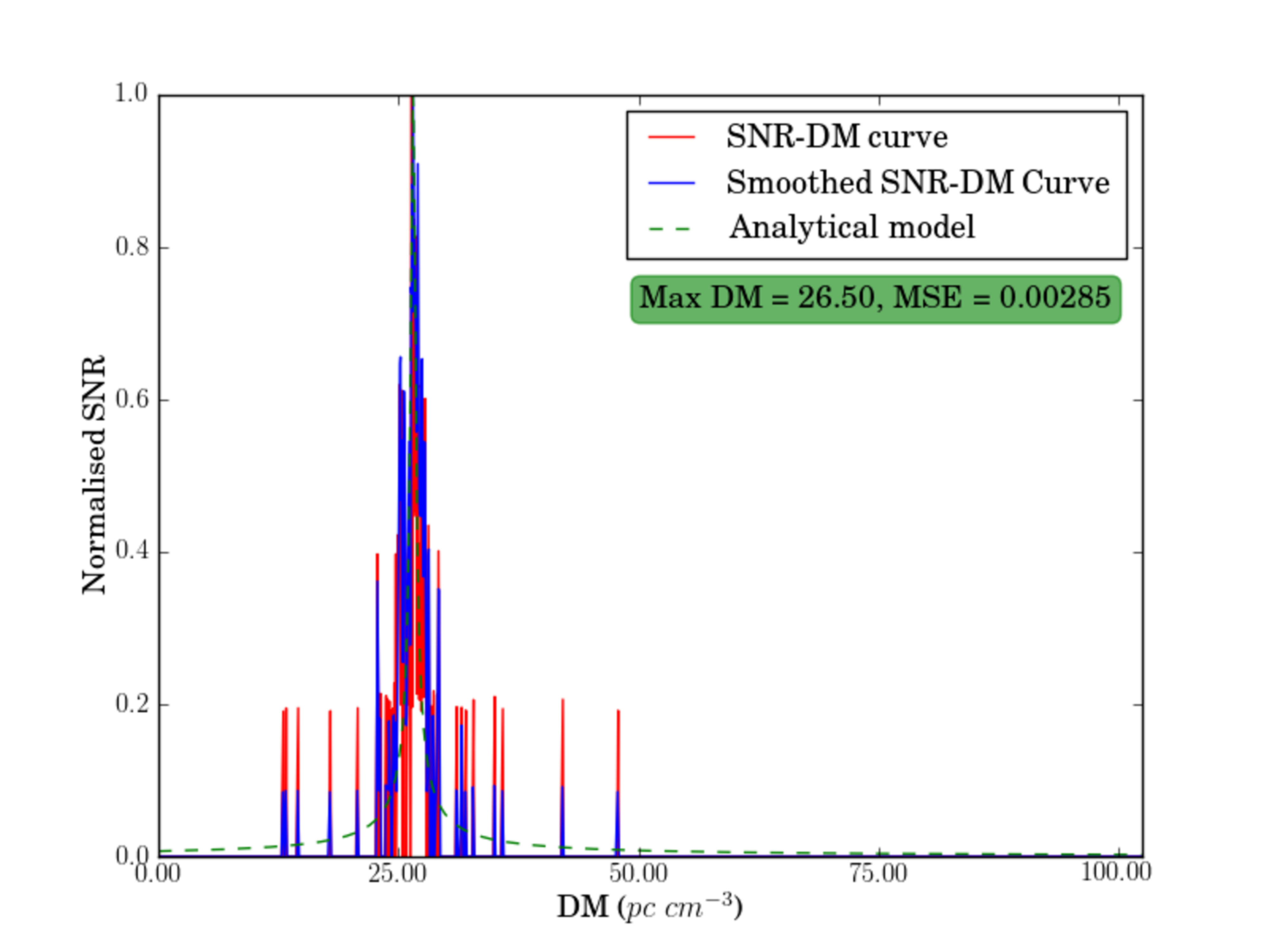}}
  \hspace{8mm}
  \subfloat[High-intensity narrowband RFI]{\includegraphics[width=220pt]{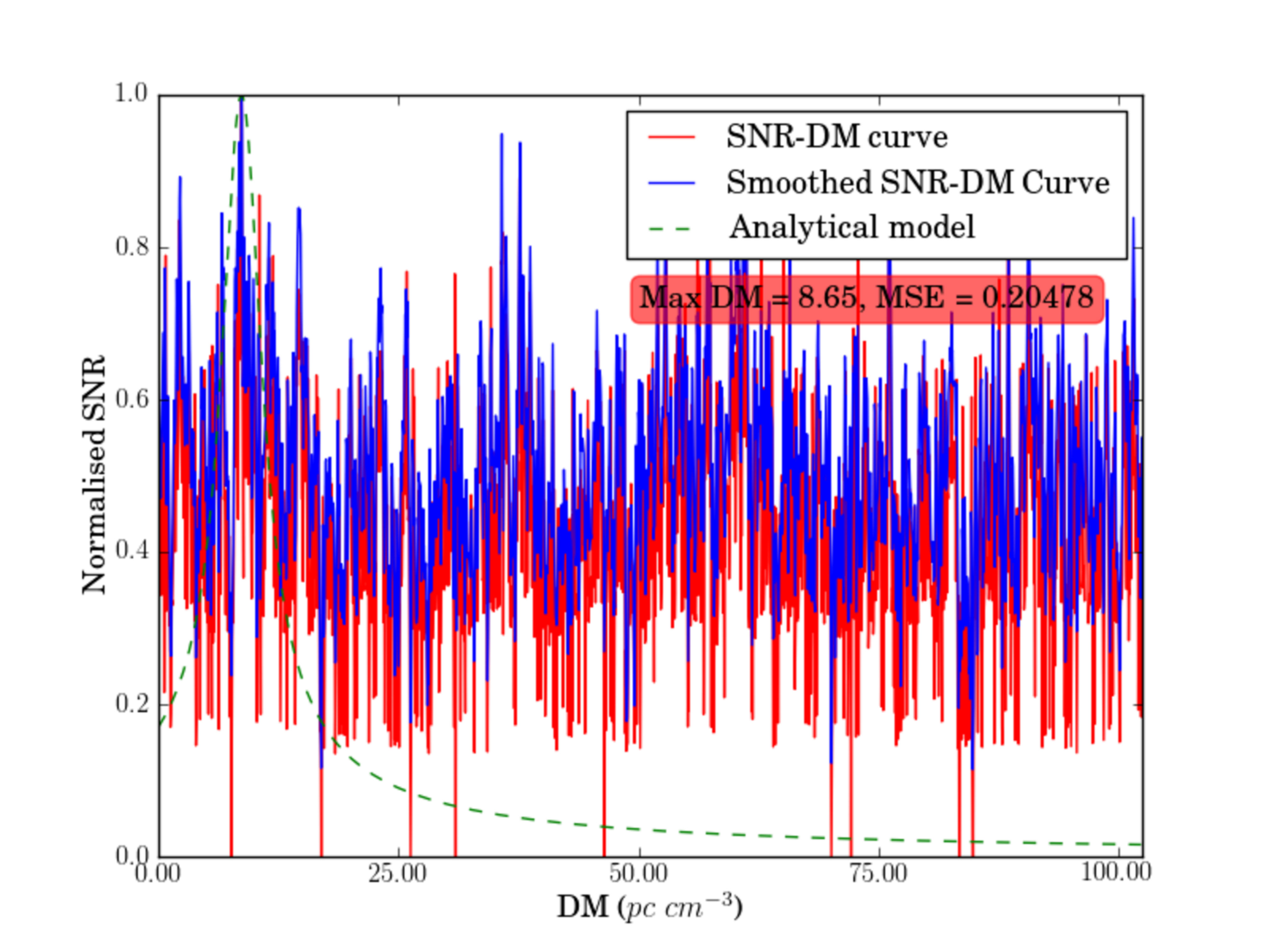}}
  \hspace{8mm}
  \subfloat[Broadband RFI]{\includegraphics[width=220pt]{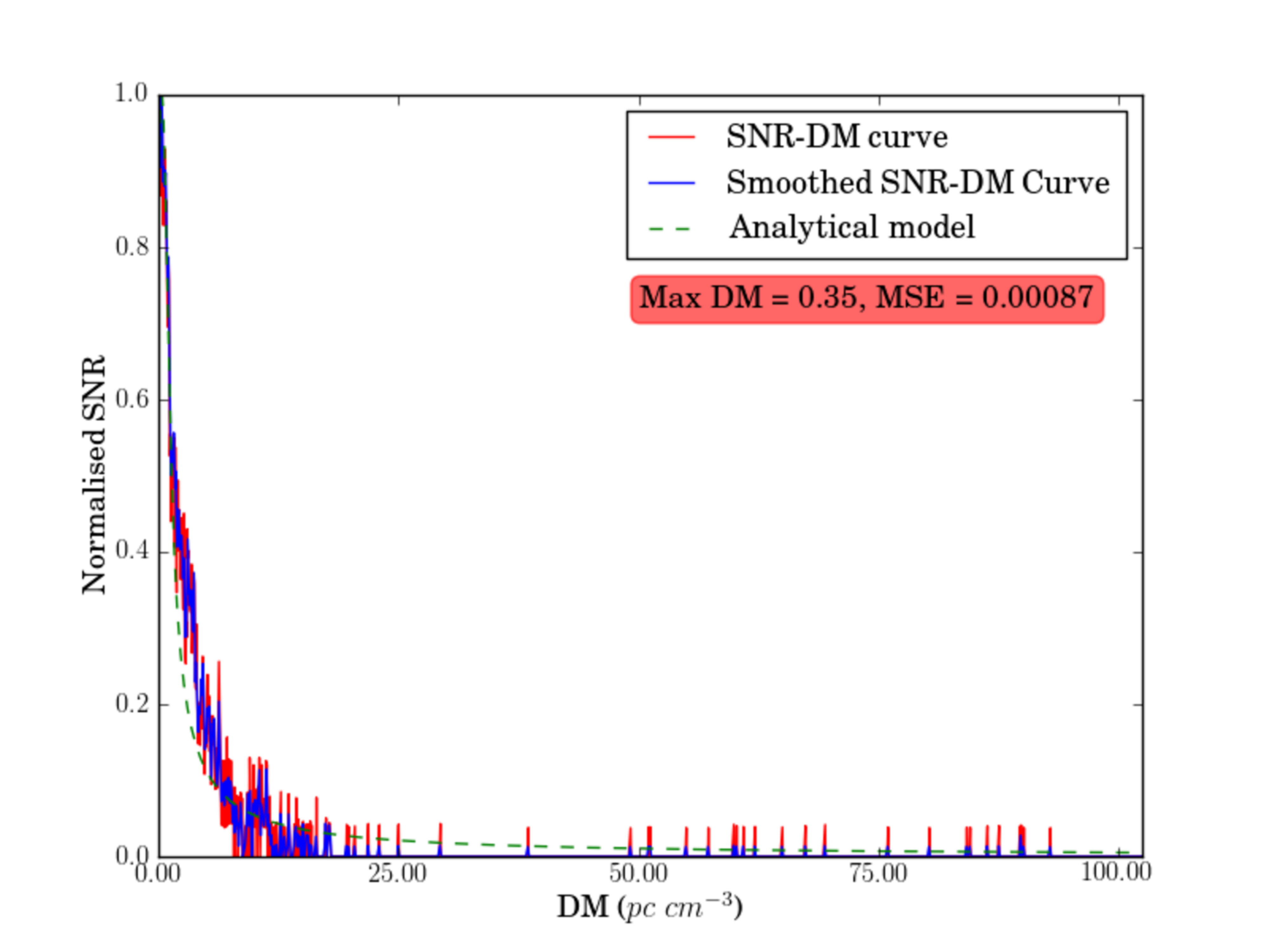}}
  \caption{Examples of clusters during the candidate selection stage. The first two plots belong to pulses from 
	  B0329+54 whilst the rest are due to RFI. Plot c represents the black vertical cluster at around 215s
	  in figure \ref{b0329Figure}, whilst plot d is the low DM, high-SNR detection at 240s.}
  \label{classificationFigure}
\end{figure}

Figure \ref{b0329Figure} represents the output of the pipeline for the previous observation for a single beam. The pulsar 
enters the beam at $\sim$100s, with pulses getting stronger as it moves towards the beam's center. Several RFI 
events were also detected, the most noticeable of them being three broadband events (the large detections at
DM $\approx$ 0) and a bright narrowband event at around 215s which was detected across the entire DM range.
Figure \ref{classificationFigure} provides visual snapshots of four clusters during the classification stage,
with two clusters originating from pulse detections from B0329+54 and two additional clusters attributed to RFI.
Both these RFI clusters were correctly filtered.

\section{Conclusion}
\label{conclusion}

We have developed a GPU-based, real-time transient detection pipeline capable of processing multiple beams 
simultaneously, with data streamed via a 10GigE link from a digital beamformer. This pipeline consists of
several processing stages, including: RFI mitigation using simple thresholding techniques, brute-force
dedispersion, density-based clustering for grouping related detections together and cluster classification
to filter out RFI-induced clusters. Interesting events, as well as the entire incoming data stream, can be
persisted to disk at any time during the pipeline's execution. A GPU server was deployed at the 
Medicina BEST-2 array near Bologna, Italy, where several benchmarks and test observations were performed, 
with the digital backend capable of streaming 8 single polarization 20MHz beams at 5.12 Gbps. With this 
setup, the pipeline could process the 8 beams on two NVIDIA GTX 660Ti cards for a maximum of 640 DM values, 
with dedispersion step and maximum DM having a negligible effect on performance. The bandwidth-limited 
dedispersion step is the slowest part of the pipeline, taking up 93\% of the GPU processing time. For 
this reason, improving this stage and implementing faster versions of the algorithm is an ongoing effort.
The current implementation 

The pipeline architecture parallelizes processing across beams, so when a single server is not capable
of processing the total number of beams output by a beamformer multiple servers can be used if the data
can be streamed to different destinations (through an intermediary switch or different interfaces on 
the backend system). Wider bandwidths require more processing and so decrease the number of beams which can
be processed. This assumes homogeneous server configuration, otherwise a load-balancing intermediary server
will be required to split the data streams. This parallelization philosophy make this prototype a viable
system for processing beamformed data as long as a single beam can be processed on one GPU, and the 
ever-increasing processing power of these devices makes this notion very realizable for future radio telescopes.

\section*{Acknowledgments}
We would like to thank Stelio Montebugnoli, Jader Monari, Germano Bianco and all the staff at the Medicina
Radio Observatory for their invaluable help on-site during deployment and observing sessions. We would also like to 
thank the system designers for the digital backend, especially Griffin Foster, for their help in designing 
the interfacing protocol between the digital backend and host system, as well as for hours of support. 

The research work disclosed in this publication is partially funded by the Strategic Educational Pathways Scholarship (Malta).

\end{document}